\documentclass[12pt]{article}
\usepackage{pdflscape}
\usepackage{xr-hyper}
\usepackage{amsmath,graphics,epsfig,lscape,setspace,amssymb,colortbl,enumerate,moresize,bm,url}
\usepackage{natbib}
\usepackage{har2nat}
\usepackage{soul}
\usepackage{multirow}
\usepackage{booktabs}
\usepackage{xr-hyper}
\usepackage{graphicx}
\usepackage[toc,page]{appendix}
\usepackage{comment}
\usepackage{etoolbox} 
\definecolor{nupurple}{RGB}{78, 42, 142}
\usepackage{amsthm}
\usepackage[dvipsnames]{xcolor}
\usepackage{placeins}
\usepackage[breaklinks=true,colorlinks=true,citecolor=nupurple]{hyperref}
\usepackage[bottom]{footmisc}
\usepackage{authblk}
\newcommand{\norm}[1]{\left\lVert#1\right\rVert}

\newcommand{\Cov}{{\rm Cov}}

\newcommand \ba  {\begin{array}}
\newcommand \ea  {\end{array}}
\newcommand \bi  {\begin{itemize}}
\newcommand \ei  {\end{itemize}}
\newcommand \ben  {\begin{enumerate}}
\newcommand \een  {\end{enumerate}}
\newcommand \be  {\begin{equation}}
\newcommand \bea {\begin{eqnarray}}
\newcommand \ee  {\end{equation}}
\newcommand \eea {\end{eqnarray}}
\newcommand{\citeN}[1]{\citeasnoun{#1}}

\newcommand{\nothing}[1]{}
\nothing{This allows you to make long comments}

\newcounter{dispsection}

\newtheorem{theorem}{Theorem}[dispsection]

\newcommand{\mysection}[2]{%
  \section{#2}%
  \setcounter{dispsection}{#1}%
  \setcounter{theorem}{0}%
}

\newtheorem{lemma}{Lemma}[dispsection]

\newtheorem{definition}[theorem]{Definition}

\newtheorem{proposition}[theorem]{Proposition}

\makeatletter
 \def\captionof#1#2{{\def\@captype{#1}#2}}
\makeatother

\newcommand{\shrinkparm}{0mm}

  {\rule[-4mm]{0mm}{6mm}\noindent\vspace{3cm}\end{table}}

  {\rule[-4mm]{0mm}{6mm}\noindent\vspace{3cm}\end{table}}

  {Table continued on next page ... \end{table}}

  {\noindent\vspace{1cm}\end{figure}}

  {\rule[-4mm]{0mm}{12mm}\noindent\rule{\linewidth}{1mm}\vspace{3cm}\end{sidewaystable}}

\makeatletter
\newcommand{\xRightarrow}[2][]{\ext@arrow 0359\Rightarrowfill@{#1}{#2}}
\makeatother

\DeclareMathOperator*{\argmin}{arg\,min}

\usepackage{titlesec}

\titleformat{\paragraph}
{\normalfont\normalsize\bfseries}{\theparagraph}{1em}{}
\titlespacing*{\paragraph}
{0pt}{3.25ex plus 1ex minus .2ex}{1.5ex plus .2ex}

\allowdisplaybreaks

\begin{document}
\onehalfspacing

\title{{\Large {\textbf{High-dimensional Penalized Linear IV Estimation \& Inference using BRIDGE and Adaptive LASSO}}}}
\author{Eleftheria Kelekidou \thanks{This research was supported in part through the computational resources and staff contributions provided for the Quest high performance computing facility at Northwestern University which is jointly supported by the Office of the Provost, the Office for Research, and Northwestern University Information Technology.}}
\affil{Northwestern University, USA}

\date{{\today}
\\[.025in]
}

\maketitle

\begin{abstract}

    This paper is an exposition of how BRIDGE and adaptive LASSO can be used in a two-stage least squares problem, to estimate the 2nd stage coefficients when the number of parameters $p$ in both stages is growing with the sample size $n$ . Facing a larger class of problems compared to the usual analysis in the literature, i.e. replacing the assumption of normal with sub-gaussian errors, I prove that both methods are model selection consistent and oracle efficient even when the number of instruments and covariates exceeds the sample size. For BRIDGE, I also prove that if the former is lower than the latter, the same properties hold without sub-gaussian errors. When $p>n$, BRIDGE requires a slightly weaker set of assumptions to have the desirable properties, as adaptive LASSO requires a good initial estimator of the relevant weights. However, adaptive LASSO is expected to be much faster computationally, so the methods are competitive on different fronts and the one that is recommended depends on the researcher's resources.
\end{abstract}

\mysection{-1}{Introduction}
\noindent 

Using increasingly large datasets, economists and social scientists in general face new challenges that are connected with the high-dimensional nature of the problems they study. A particular instance of such a scenario is the class of applications that involve high-dimensional instruments. Valid instruments are often the holy grail of applied research; however, recently there is increased availability in rich datasets such as the GWAS Catalog \cite{Macarthur:2017, Sollis:2023}, which includes information on how phenotypic variations affect biological traits and disease traits give new paths for research. This type of variables can provide a number of instrument alternatives for applications related to productivity, risk, welfare, actuarial science, etc \cite{Gui:2005, Ma:2007, Wang:2008}. In order to be able to do estimation and inference in environments like this, the researcher has to address the statistical difficulties that occur.

 I will focus on the usual linear model \[Y=X'\beta_0 +u\] where the usual exogeneity restriction, i.e. $E[u|X]= 0$ fails. Instead, I will assume that there exist observable variables $Z$ that can be used as instruments, such that $E[u|Z]= 0$, $\Cov[ZX]$ is full rank, and the conditional expectation $E[X|Z]$ is linear. I am allowing the number of both the endogenous covariates and the instruments to grow and I am making the high level assumption that the number of instruments is sufficient for the identification of the model. 
 
 Due to the dimensionality problem, if the number of covariates in each stage exceeds the sample size, the OLS estimator is not identified. In this case, one can obtain valid estimators by minimizing the penalized loss function, i.e. \(L_n(b)= \argmin_b \frac{1}{2}\sum_{i=1}^{n}(Y_i-X_i'b)^2 +\lambda_n\sum_{j}^{p}p(b_j)\) where $p(b_j)$ is a penalty function. Due to the presence of endogeneity, instead of $X_i$, I will use an estimator of $E[X|Z]$. Under sparsity, the latter can be obtained by using any proper penalization function. Then, I will estimate the model choosing the following popular penalty functions: BRIDGE , with $p(b)=|b|^{\gamma}$, where $\gamma\in (0,1)$ and adaptive LASSO with $p(b)=w_b|b|$, where $w_b$ is a set of weights. 

 \textbf{Related literature:} The standard practice to estimate high-dimensional, sparse, linear models includes popular penalization functions, such as, for $\gamma=1$, LASSO \cite{Tibshirani:1996}, adaptive LASSO \cite{Zou:2006}, BRIDGE \cite{Knight:2000}, the smoothly clipped absolute deviation (SCAD) \cite{Fan:2001}, the minimax concave pentaly (MCP) \cite{Zhang:2010} etc. However, inference in these setups often remains a difficult problem even with the sole existence of exogenous covariates \cite{Van:2014}.  In environments that endogeneity is also an issue, the econometrician faces three distinct scenarios; the first is having a growing number of instruments, yet a small number of covariates of interest \cite{Belloni:2012,Caner:2015, Fan:2018, Hansen:2014}. The second is allowing for a growing number of covariates but a low-dimensional set of them being endogenous, which permits for a low dimensional estimation in the first stage (eg. \citeN{Fan:2014}). Finally, there is the case with high-dimensionality in both stages with a growing number of instruments and regressors. \citeN{Lin:2015} allow for this setting and they prove model selection consistency and oracle efficiency using LASSO, SCAD, and MCP, under the restrictive assumption of normality. \citeN{Caner:2014} propose a GMM estimator using adaptive elastic nets as a penalty function, allowing for a growing number of covariates but without exceeding the sample size.\citeN{Gold:2020} follows a similar direction, providing general assumptions under which a penalty function would have the desirable properties but restricting the nature of the instruments to sub-Gaussian. Also, for LASSO, the required properties imply the presence of strong compatibility conditions to achieve the oracle property.  
 
To address the need for a more flexible framework, a natural path would be to turn to a penalty function that ensures model selection consistency with minimal assumptions in the simple OLS case, and consider the extension to the endogeneity framework. BRIDGE and adaptive LASSO are natural candidates for this setup.  There have been some attempts \cite{Bahador:2024} to apply BRIDGE in two-stage penalized least squares using a control function approach, yet there is no formal proof of model selection and they require normality for the error terms. 

\textbf{Contribution:} The main contribution of this paper is showing that the estimator for $\beta_0$ using both penalty functions is model selection consistent and oracle efficient. For both methods, if the number of parameters grows faster than the sample size, the result is based only on sub-Gaussian tails of the error term and standard rate conditions. The argument for BRIDGE is an extension to the partial orthogonality condition in \citeN{Huang:2008}. For adaptive LASSO, the proof is based on the requirement of a consistent estimator for the weights in the penalty function as stated for the exogenous case in \citeN{Huang:2008b}. Moreover,  I show that for a growing number of parameters but not higher than the sample size, the result for BRIDGE holds with just homoskedastic, mean 0 errors and the corresponding rate assumptions. Lastly, for BRIDGE,  I provide a proposal of a consistent estimator for the standard errors as well as some computational evidence on the performance of the method. Corresponding exercises on adaptive LASSO are ongoing and available upon request.

\textbf{Structure:} The rest of the paper is organized as follows. Section 2 introduces the notation and the requirements for the 1st stage of the problem, regardless of the 2nd stage method. In Section 3, I present the results for BRIDGE when $p<n$. In Sections 4 and 5, I present the results for $p>n$ for BRIDGE and adaptive LASSO correspondingly. Finally, Section 6 contains some computational results comparing how BRIDGE and LASSO perform in this problem.


\mysection{0}{Notation \& Setup}

\noindent 

Consider the vector $(Y, X,Z )$, where $Y$ is the outcome random variable, $X$ is the random vector of observable covariates - possibly endogenous, and $Z$ is a fixed vector of instruments. The researcher observes a random sample of size $n$. Assume that the true conditional expectation $E[X|Z]$ is linear.

This paper is concerned with the following two-stage model: 
\begin{align*}
     1^{st} stage: \quad& X_{ij} = Z_i'\alpha_j + v_{ij},\  for\ i=1,\dots,n\ \&\ j=1,\dots,p_{xn}\\
     2^{nd}   stage: \quad& Y_i=X_i'\beta_0 +u_i,\ \ \ for\ i=1,\dots,n
\end{align*}

where $X_i=(X_{i1}, \dots, X_{ip_{xn}})'\in \mathbb{R}^{p_{xn}}$ consists of the covariates in the $2^{nd}$ stage with $\beta_0\in \mathbb{R}^{p_{xn}}$ being the corresponding coefficients, and $Z_i\in \mathbb{R}^{p_{zn}}$ consists of the covariates in the $1^{st}$ stage with $\alpha_j\in \mathbb{R}^{p_{zn}}$ being the corresponding coefficients.  Let $\alpha\in \mathbb{R}^{p_{zn}\times p_{xn}}$ the matrix of the stacked first stage coefficients. I define the respective conditional means as $d_{ij}:=E[X_{ij}|Z_i]=Z_i'\alpha_j $. Let the matrix of the second stage covariates be $X_n\in \mathbb{R}^{n\times p_{xn}}$. Accordingly, let the matrix of the first stage covariates be $Z_n\in \mathbb{R}^{n\times p_{zn}}$.  Further,  I define $D_n=E[X_n|Z_n]=Z\alpha\in \mathbb{R}^{n\times p_{xn}}$ , which rows are $d_i=(d_{i1}, \dots, d_{ip_x})'$.

Naturally, $u_i\ and\ \bm{v}_{i}:= (v_{i1}, v_{i2},\dots, v_{ip_x})'$ are the random noise terms in each stage, and satisfy $E[u_i|Z_i]=0,\ E[v_i|Z_i]=0$ but not necessarily $E[u_i|X_{ij}]=0$. Finally, note that the model above implies that $Y_i= (Z_i'\alpha)'\beta_0+v_i'\beta_0+u_i= d_i'\beta_0+\underbrace{v_i'\beta_0+u_i}$, so I define $\varepsilon_{i}=v_i'\beta_0+u$ the new error term that satisfies $E[\varepsilon_i|Z_{i}]=0$.

Define $X_{1n}\in\mathbb{R}^{n\times k_{xn}}$ the matrix of the relevant covariates in the 2nd stage and $X_{2n}\in\mathbb{R}^{n\times m_{xn}}$ the matrix of the rest. Let $\beta_{01}$ be the sub-vector of 2nd stage coefficients that are non $0$, and $\beta_{02}$ vector of the rest of elements of $\beta_0$. Similarly for the first stage; $Z_{1nj}\in \mathbb{R}^{n\times k_{znj}}$ is the matrix of relevant covariates corresponding to the 2nd stage covariate $j$ and $Z_{2nj}\in\mathbb{R}^{n\times m_{knj}}$ the matrix of the rest. Let $\alpha_{1j}$ be the sub-vector of the non-zero coefficients of $\alpha_j$ $\forall j$,  and $\alpha_{2j}$ the sub-vector of the zero coefficients.  The number of instruments and the number covariates are allowed to grow with the sample size but the model should be sparse in both stages; that is, $k_{xn}/n\to 0 $ and $\max k_{znj}/n\to 0$.

For simplicity of notation, let $Y_i$ be centered and the instruments be standardized, ie.

\[
\sum_{i=1}^{n}Y_i=0, \quad \sum_{i=1}^{n}Z_{ih}=0, \quad \frac{1}{n}\sum_{i=1}^{n}Z_{ih}^2=1,
\]
for all instruments $h=1,\dots,p_{zn}$.

Lastly, for a given vector $\delta$, $\norm{\delta}$ is the Euclidean norm, and for a given matrix $\Delta$, $\norm{\Delta}$ is the spectral norm.

\subsection*{1st Stage}

Due to existing results, the 1st stage coefficients are easy to deal with. For their estimation, the researcher can use the method of choice that provides model selection consistency and oracle efficiency. However, the assumptions that are needed in order to use the said method will directly affect the distribution of the 2nd stage BRIDGE coefficients. Define $\Sigma_{zn}=n^{-1}Z_n'Z_n$. Let $\rho_{1n}^z$ be the smallest and $\rho_{2n}^{z}$ the largest eigenvalue of $\Sigma_{zn}$. Define $\Sigma_{1nj}=n^{-1}Z_{1nj}'Z_{1nj}$. Let $\tau_{1n}^z$ be the smallest and $\tau_{2n}^{z}$ the largest eigenvalue of $\Sigma_{1nj}$.     A sufficient set of conditions for the 1st stage, in order to achieve the desirable properties in the 2nd stage, are the following:

\subsubsection*{Assumptions}
For each $j$:
\begin{enumerate}
    \item[\textbf{A.1}] $v_{1j},\dots,v_{nj}$ are i.i.d. random variables with mean $0$ and variance $\sigma_j^2,\ 0<\sigma_j^2<\infty,\ \forall j=1,\dots,p_{xn}$. Also, $Cov(v_{ij},v_{ih})=\sigma_{v_j,v_h}<\infty,\ \forall j,h$.
    \item[\textbf{A.2}]  (a) $\exists \text{ constant }0<\rho_2^z<\infty$ , st. $0<\rho_{1n}^z<\rho_{2n}^z<\rho_2^z<\infty$ for all $n$. (b) $\exists$ constants $0<\tau_{1z}<\tau_{2z}<\infty$ st $\tau_{1z}<\tau_{1nj}<\tau_{2nj}<\tau_{2z},\ \forall n$. 
    \item[\textbf{A.3}] $\exists$ constants $0<a_0<a_1<\infty$, st. $a_0\leq min\{|\alpha_{1hj}|,1\leq h\leq k_{znj}\}\leq max\{|\alpha_{1hj}|,1\leq h\leq k_{znj}\}\leq a_1$.
    \item[\textbf{A.4}] The estimator $\hat{\alpha}$ used is model selection consistent and oracle efficient.

\end{enumerate}

(A1) ensures the good behavior of the composite error term and allows for a complicated dependence structure between the regressors. The second assumption ensures the good behavior of the weighted Gram matrices $\Sigma_{zn},\ \Sigma_{1nj}$. Assumption (A3) ensures that the small coefficients are far enough from 0 and the large ones do not diverge. Finally, (A4) is the most high level assumption, ensuring knowing the limit distribution of $\hat{\alpha}$. Since this is a regular high-dimensional setting, this is not a difficult assumption to satisfy and it can be replaced with the set of assumptions of a specific choice of penalty function. For example, for BRIDGE, (A1-A3) and a set of rate assumption is sufficient for (A4) \cite{Huang:2008}. For fewer instruments than the sample size, the assumption on the error term will be exactly the same, while for $p_{zn}>n$ the errors of the 1st stage should also be sub-Gaussian to satisfy the conditions of the aforementioned paper.


\mysection{1}{BRIDGE for $p<n$}

Having ensured a good estimator for the first stage, I only need to provide an estimator with good properties for the second stage. I first state the results for $p_{xn}<n$. In the case of $p_{xn}>n$, these results will be used after ensuring model selection consistency. That is, after picking the correct model, due to sparsity, I can apply the same results as in $p_{xn}<n$, and ensure consistency and oracle efficiency even in the most cumbersome case.

Let $\hat{\beta}_j\in\mathbb{R}^{p_{xn}}$ be the root of the following minimization problem: \[\hat{\beta}=\argmin_{b}\underbrace{ \frac{1}{2}\sum_{i=1}^{n}(Y_i-\hat d_i'b)^2+\lambda_{xn}\sum_{j=1}^{p_{xn}}|b_{j}|^\gamma}_{L_n(b)}.\]

where $\lambda_{xn}$ is the tuning parameter. Define $\Sigma_{dn}=n^{-1}D_n'D_n$. Let $\rho_{1n}^d$ be the smallest and $\rho_{2n}^{d}$ the largest eigenvalue of $\Sigma_{dn}$. Let $D_{1n}\in\mathbb{R}^{n\times k_{xn}}$ be the matrix of the conditional expectations of the relevant covariates in the 2nd stage and $D_{2n}\in\mathbb{R}^{n\times m_{xn}}$ the matrix of the rest. Define $\Sigma_{1dn}=n^{-1}D_{1n}'D_{1n}$. Let $\tau_{1n}^d$ be the smallest and $\tau_{2n}^{d}$ the largest eigenvalue of $\Sigma_{1dn}$.

Also, define $\hat \Sigma_{dn}=n^{-1}\hat D_n'\hat D_n$. Let $\hat\rho_{1n}^d$ be the smallest and $\hat\rho_{2n}^{d}$ the largest eigenvalue of $\hat\Sigma_{dn}$. Let $\hat D_{1n}\in\mathbb{R}^{n\times k_{xn}}$ the matrix of the estimated conditional expectations of the relevant covariates in the 2nd stage and $\hat D_{2n}\in\mathbb{R}^{n\times m_{xn}}$ the matrix of the rest. Define $\hat \Sigma_{1dn}=n^{-1}\hat D_{1n}'\hat D_{1n}$. Let $\hat\tau_{1n}^d$ be the smallest and $\hat\tau_{2n}^{d}$ the largest eigenvalue of $\hat\Sigma_{1dn}$. Let $\hat d_{1i}:=Z_i'\hat\alpha $ a row of $\hat D_{1n}$ and the rest of the vectors accordingly.

  \subsubsection*{Assumptions}
\begin{enumerate}
    \item[\textbf{B.1}] $u_{1},\dots,u_{n}$ are i.i.d. random variables with mean $0$ and variance $\sigma_u^2,\ 0<\sigma_u^2<\infty$. Also, $Cov(u_i, v_{ij})=\sigma_{u,v_j},\ 0<\sigma_{u,v_{j}}<\infty,\ \forall j$.
    \item[\textbf{B.2}]  $\exists \text{ constants }0<\rho_1^d<\rho_2^d<\infty$ , st. $0<\rho_1^d<\rho_{1n}^d<\rho_{2n}^d<\rho_2^d<\infty$ for all $n$. 
    \item[\textbf{B.3}] (a) $\lambda_{xn}(k_{xn}/n)^{1/2}\to 0$. (b) $(\lambda_{xn}k_{xn}+p_{xn})/n\to 0$.  (c) $p_{xn}^2\max_jk_{znj}/n\to0$.  (d) $k_{xn}/n^{1/2}\to 0$. (e) $p_{xn}/n\to 0$. (f) $\lambda_{xn}n^{-\gamma/2}/(\sqrt{p_{xn}k_{xn}\max_jk_{znj}})^{2-\gamma}\to\infty$.
    \item[\textbf{B.4}] $\exists$ constants $0<b_0<b_1<\infty$, st. $b_0\leq min\{|\beta_{01j}|,1\leq j\leq k_{xn}\}\leq max\{|\beta_{01j}|,1\leq j\leq k_{xn}\}\leq b_1$.
    \item[\textbf{B.5}] (a) $\exists$ constants $0<\tau_{1}<\tau_{2}<\infty$ st $\tau_{1}<\tau_{1n}^d<\tau_{2n}^d<\tau_{2},\ \forall n$. (b) $n^{-1/2}\max_{1\leq i\leq n}d_{1i}'d_{1i}\to 0$. 
    \item[\textbf{B.6}] The instruments are $O_p(1)$.
\end{enumerate}

Assumption (B.1) is of the same nature as (A.1). The 2nd stage error term is homoskedastic and the covariance with each one of the 1st stage error terms needs to be finite but is allowed to be non-zero. Combining the expressions of the two stages, the error term of interest is $\varepsilon_i=v_{i}'\beta_0+u_i$ which is i.i.d. with mean 0 and variance $\sigma_{\varepsilon}^2$. Assumption (B.2) ensures the invertibility of $\Sigma_{dn}$ but, contrary to the smallest eigenvalue of $\Sigma_{zn}$, $\rho_{1n}^{d}$ is not allowed to converge to 0. A direct consequence is that the convergence rates of the 2nd stage estimator will no longer depend on the eigenvalues of the corresponding Gram matrix in the way the 1st stage ones do. (B.3.b) is used for the convergence of the 2nd stage estimator. (B.3.b) ensures model selection consistency and the rest of the rate assumptions are used for the three upcoming results. Note that the rates are affected by the number of the instruments and the maximum number of relevant coefficients of the 1st stage separate equations. Assumption (B.4) is the same as (A.4) and common in high dimensional literature, as it ensures that the relevant coefficients are uniformly bounded away from 0 (and infinity). Assumption (B.5.a) is uniformly bounding the eigenvalues of $\Sigma_{1n}^d$ from $0$ and infinity, making it strictly positive definite. Assumption (B.5.b) is used in the asymptotic normality proof and is implied by (B.3.e) if the conditional expectations of the relevant covariates in the 2nd stage are uniformly bounded. (B.6) is used in the asymptotic normality proof.

Under the set of assumptions above, and the identification assumption that there are sufficiently many instruments for the endogenous covariates, I prove the following two results, the former for regarding the consistency of the 2nd stage BRIDGE estimator and the second regarding the model selection consistency and oracle efficiency that it achieves.

\begin{theorem}\label{thm:consistency}
    Under (A.1-A.4) and (B.1-B.3a-d,B.4), let $h_n=\frac{(p_{xn}k_{xn}\max_jk_{znj})^{1/2}}{\sqrt{n}}$ and $h_n'=\left(\frac{\lambda_{xn}k_{xn}+p_{xn}}{n}\right)^{1/2}$. Then, $\norm{\hat{\beta}_n-\beta_0}=O_p(\min\{h_n,h_n'\})$.
\end{theorem}

     Note that \ref{thm:consistency} utilizes all (A.1-A.4) as the oracle efficiency of the first stage coefficients is used in the proof to determine the rate of $\norm{\hat{\alpha}-\alpha}$. Note that neither rate is affected by the actual number of the instruments but only on the maximum number of the relevant instruments among the endogenous regressors. It is not straightforward which of the rates is faster as, apart from $p_{xn}, k_{xn}$ and the sample size, they depend on different quantities. That is, $h_n$ also depends on the maximum number of relevant covariates in the separate 1st stage equations, while $h_n'$ depends on the tuning parameter $\lambda_{xn}$. Interestingly, the latter ensures consistency and is used as a middle step to prove the former rate which is essential for model selection consistency.

\begin{theorem}\label{thm:oracle}
    Let $\hat{\beta}_n=(\hat{\beta}_{1n},\hat{\beta}_{2n})$ where $\hat{\beta}_{1n},\hat{\beta}_{2n}$ are estimators for $\beta_{01},\beta_{02}$, respectively. Suppose that $0<\gamma<1$ and that conditions (A.1-A.5) and (B.1-B.5) are satisfied. Then, \begin{enumerate}[i)]
        \item $\hat{\beta}_{2n}=0$ with probability converging to 1.
        \item Let $s_{dn}^2=\sigma_{\varepsilon}^2\delta_n'(\Sigma_{1n}^{d})\delta_n$ for all $(k_{xn}\times1)$ vectors $\delta_n$, such that $\norm{\delta_n}\leq1$. Then, $n^{1/2}s_{dn}^{-1}\delta_n'(\hat{\beta}_{1n}-\beta_{01})=n^{-1/2}s_{dn}^{-1}\sum_{i=1}^{n}\varepsilon_i\delta_n'(\Sigma_{1n}^{d})^{-1}\hat{d}_{1i}+o_p(1)\xrightarrow{d}N(0,1)$ where $o_p(1)$ is a term that converges to 0 in probability uniformly with respect to $\delta_n$.
    \end{enumerate}
\end{theorem}

The first part of this result states that the method described gives exact zero estimated coefficient values for the sub-vector of $\hat{\beta}_n$ that corresponds to the irrelevant covariates. The second part, written in a similar fashion as the result in \cite{Huang:2008}, states that the estimated non-zero coefficients are oracle efficient- that is, they asymptotically have the same distribution as they would if they were ex ante known and being the only ones used in the model. The result is written for the linear combination of the corresponding estimated sub-vector and it is straightforward to rewrite it for the marginal distributions. For each $j=1,\dots,k_{xn}$, I can pick $\delta_n= e_j$ where $e_j$ the $k_{xn}\times 1$ unit vector with $0$ elements everywhere apart from the $j$-th position. Also, define $s_{dnj}^2=\sigma_{\varepsilon}^2e_j'(\Sigma_{1n}^{d})e_j$. Then, for each $\hat{\beta}_{1nj}$, it holds that $n^{1/2}s_{dnj}^{-1}(\hat{\beta}_{1nj}-\beta_{01j})\xrightarrow{d}N(0,1)$.


\mysection{2}{BRIDGE for $p>n$}

Keeping (A1-A4) and adding some more structure to the error term, the following list of assumptions is sufficient for model selection consistency and oracle efficiency under the BRIDGE penalty function.

\subsubsection*{Assumptions}
\begin{enumerate}
    \item[\textbf{C.1}] (a) $u_{1},\dots,u_{n}$ are i.i.d. random variables with mean $0$ and variance $\sigma_j^2,\ 0<\sigma_u^2<\infty$. Also, $Cov(v_{ij},v_{ih})=\sigma_{v_j,v_h}<\infty,\ \forall j,h$. (b) $u_i,\{v_{ij}\}_{j=1}^{p_{xn}}$ are jointly sub-Gaussian; let $\vec{u}= \{u_i,\{v_{ij}\}_{j=1}^{p_{xn}}\}$, then $\norm{\vec{u}}_{\psi_2}=\sup_{\norm{a}_2\leq1}\norm{a'\vec{u}}_{\psi_2}\leq \infty$.
    \item[\textbf{C.2}] Partial orthogonality: (a) $\exists$ constant $c_{0}$ st. $|n^{-1/2}\sum_{i=1}^{n}d_{ij}d_{ik}|\leq c_0$, $j=1,\dots,m_{xn},\ k\in 1,\dots,k_{xn}$. (b) Define $\xi_{nj}=n^{-1}E(\sum_{i=1}^{n}Y_{i}d_{ij})=n^{-1}\sum_{i=1}^{n}d_{1i}'\beta_{10}d_{ij}$, then there exists a constant $\xi_0>0$ st. $\min_{k\in k_{xn}}|\xi_{nk}|>\xi_0>0$.
    \item[\textbf{C.3}] $\lambda_{xn}/n\to 0$ \& $\lambda_{xn}n^{-\gamma/2}k_{xn}^{\gamma/2}\to\infty$; (b) $log(m_{xn})=o(1)(\lambda_{xn}n^{-\gamma/2})^{2/(2-\gamma)}$
    \item[\textbf{C.4}]  $\exists$ constants $0<b_0<b_1<\infty$, st. $b_0\leq min\{|\beta_{1k}|,1\leq k\leq k_{xn}\}\leq max\{|\beta_{1k}|,1\leq k\leq k_{xn}\}\leq b_1$.
    \item[\textbf{C.5}] (a) $\exists$ constants $0<\tau_{1}<\tau_{2}<\infty$ st $\tau_{1z}<\tau_{1n}<\tau_{2n}<\tau_{2},\ \forall n$. (b) $n^{-1/2}\max_{1\leq i\leq n}d_{1i}'d_{1i}\to 0$. 
    \item[\textbf{C.6}] (a) $k_{xn}(1+\lambda_{nj}^*)/n\to 0 $ \& $\lambda_{nj}^*(k_{xn}/n)^{1/2}\to 0$ where $\lambda_{nj}^*$ is the tuning parameter from the partial optimization problem $U_n^*(\beta_{10})$ (see Appendix for details). (b) $k_{xn}^2\max_jk_{znj}/n$.
    \item[\textbf{C.7}] The instruments are $O_p(1)$.
\end{enumerate}

 (C.1.a) is the same as in the previous case, while (b) relaxes the standard assumption of normality that is usual in this problem and is used to prove model selection consistency. (C.2) follows \citeN{Huang:2008} and limits the correlation between relevant and irrelevant variables and ensures that the relevant ones are sufficiently important in the model. This is also a crucial element to prove selection consistency. For the same proof, I use the rate restrictions in (C.3); it is interesting to see that there is no direct restriction on the total number of the instruments, or the total number of covariates. (C.4) is the same as for BRIDGE when $p<n$, just avoiding the restriction on the eigenvalues of the full Gram matrix. (C.5) and (C.6) are used for the oracle efficiency proof, with the former used for the Lindeberg condition. The latter includes rate restrictions to ensure consistency and oracle efficiency. Since I have ensured model selection, I can now use the previous results for $p<n$, but using $k_{xn}$ as the number of covariates. Thus, these rates look very similar with the ones used in the previous section for the same purpose, by substituting $k_{xn}$ for $p_{xn}$. The following two theorems state the results formally:

\begin{theorem}
\label{thm:large_msc}
 Under (C.1-C.4) and $0<\gamma<1$, $P(\hat{\beta}_{2n}=0)\to 1$ and $P(\hat{\beta}_{1n}\neq0, k \in k_{xn})\to 1$.
\end{theorem}

\begin{theorem}
\label{thm:large_oracle_bridge}
     Under (C.1-C.6) and $0<\gamma<1$.  Let $s_{n}^2=\sigma_{\varepsilon}^2\delta_n'(\Sigma_{1n}^{d})^{-1}\delta_n$ for all vectors $\delta_n $ of size $(k_{xn}\times1)$ st. $\norm{\delta_n}\leq1$. Then, $n^{1/2}s_{n}^{-1}\delta_n' (\hat{\beta}_{1n}-\beta_{10})=n^{-1/2}s_{n}^{-1}\sum_{i=1}^{n}\varepsilon_{i}\delta_n'(\Sigma_{1n}^{d})^{-1}d_{1i}+o_p(1)\xrightarrow{d}N(0,1)$.

\end{theorem}


\mysection{3}{Adaptive LASSO for $p>n$}

In this section, the same notation will hold as well as the normalization of the variables. I now define the adaptive LASSO problem. I assume that an initial estimator $\tilde\beta_n$ is available and I define the weights and the relevant loss function: 
\begin{align*}
    w_{nj}&=|\tilde\beta_{nj}|^{-1},\ j=1,\dots,p_{xn}\\
    L_n(b)&= \frac{1}{2}\sum_{i=1}^{n}(Y_i-\hat d_i'b)^2+\lambda_{xn}\sum_{j=1}^{p_{xn}}w_{nj}|b_{j}|
\end{align*}

Also, let $\Sigma_{n12}^d=(\Sigma_{n21}^{d})' =n^{-1}D_{1n}'D_{2n}$, as well as $H_n=D_{1n}(D_{1n'}D_{1n})^{-1}D_{1n}'$. Furthermore, for a vector $\delta$, let $sgn(\delta)=(sgn(\delta_1),sgn(\delta_2),\dots)$, and $\hat{\delta}=_s\delta$ if and only if $sgn(\hat{\delta})=sgn(\delta)$.

Continuing to assume (A1-A4) and adding some more structure to the error term, the following list of assumptions is sufficient for model selection consistency and oracle efficiency under the Adaptive LASSO penalty function.

\subsubsection*{Assumptions}
\begin{enumerate}
    \item[\textbf{D.1}] (a) $u_{1},\dots,u_{n}$ are i.i.d. random variables with mean $0$ and variance $\sigma_j^2,\ 0<\sigma_u^2<\infty$. Also, $Cov(v_{ij},v_{ih})=\sigma_{v_j,v_h}<\infty,\ \forall j,h$. (b) $u_i,\{v_{ij}\}_{j=1}^{p_{xn}}$ are jointly sub-Gaussian; then, for the tail probabilities of $\varepsilon_i$ holds that $P(|\varepsilon_i|>x)\leq K\exp(-Cx^d)$, for constants $C,K$.
    \item[\textbf{D.2}] The initial estimator $\tilde{\beta}_n$ is zero-consistent$^*$ with rate $r_n\to \infty$.
    \item[\textbf{D.3}]  $\exists$ constants $0<b_0<b_1<\infty$, st. $b_0\leq min\{|\beta_{1k}|,1\leq k\leq k_{xn}\}\leq max\{|\beta_{1k}|,1\leq k\leq k_{xn}\}\leq b_1$.
    \item[\textbf{D.4}] (a) $\lambda_n\to\infty$, (b) $\lambda_{nx}k_{xn}/n^{1/2}\to 0 $, (c) $p_{xn}^2(k_{xn}\max_jk_{znj}/n)\to 0$\\
    (d) for $1<d\leq 2$, \[\frac{(\log k_{xn})^{1/d}}{\sqrt{n}}\to 0,\text{ and } \frac{\lambda_{xn}k_{xn}}{n}\to 0\]\[\frac{\sqrt{n}(\log m_{xn})^{1/d}}{\lambda_{xn}r_n}\to 0,\text{ and } \frac{k_{xn}^2}{r_n}\to 0\]
    (e) for $d=1$, \[\frac{(\log n)(\log k_{xn})}{\sqrt{n}}\to 0,\text{ and } \frac{\lambda_{xn}k_{xn}}{n}\to 0\]\[\frac{\sqrt{n}(\log n)(\log m_{xn})}{\lambda_{xn}r_n}\to 0,\text{ and } \frac{k_{xn}^2}{r_n}\to 0\]
    \item[\textbf{D.5}] (a) $\exists$ constants $0<\tau_{1}<\tau_{2}<\infty$ st $\tau_{1z}<\tau_{1n}<\tau_{2n}<\tau_{2},\ \forall n$. (b) $n^{-1/2}\max_{1\leq i\leq n}d_{1i}'d_{1i}\to 0$. (b) $\exists \text{ constants }0<\rho_1^d<\rho_2^d<\infty$ , st. $0<\rho_1^d<\rho_{1n}^d<\rho_{2n}^d<\rho_2^d<\infty$ for all $n$. (c) $n^{-1/2}\max_{1\leq i\leq n}d_{1i}'d_{1i}\to 0$. 
    \item[\textbf{D.6}] The instruments are $O_p(1)$.

\end{enumerate}

where for (D.2) I use the following definition from the online appendix of \citeN{Huang:2008b}:

\begin{definition}
    The initial estimator $\tilde\beta_n$ is 0-consistent if and only if (a) $\max_{1\leq j\leq m_{xn}}|\tilde\beta_{nj}|=o_p(1)$ and (b) there exists a constant $\xi_b>0$ such that, for any $\epsilon>0$, \[P\left(\min_{1\leq j \leq k_{xn}}|\tilde\beta_{nj}|\geq\xi_b b_1\right)>1-\epsilon\] for $n$ sufficiently large. In addition, $\tilde{\beta}_n$ is a zero-consistent with rate $r_n$ if (a) is strengthened to \[r_n\max_{1\leq j\leq m_{xn}}|\tilde\beta_{nj}|=O_p(1),\] where $r_n\to\infty$.
\end{definition}

(D.1) is crucial for the proof of model selection consistency, as it allows for the use of the relevant maximal inequalities. Assumption (D.2) states the strong prerequisite for the success of adaptive lasso; having a good initial estimator to calculate the weights. This can be weakened slightly by assuming a weaker consistency notion and adding a partial orthogonality condition in the same style as in BRIDGE. The exact condition is discussed formally in \citeN{Huang:2008b}. (D.3) bounds the parameters that correspond to the relevant regressors away from 0 and away from drifting to infinity. The former is a stronger requirement as it does not allow for positive but arbitrarily small values. (D.4) includes all relevant rates conditions necessary. While the first three are used in multiple occasions, (d) and (e) are paired with the corresponding maximal inequalities that explore the $\psi_d$ norm of a given sub-gaussian term. (D.5) bounds the eigenvalues of the Gram matrix and the relevant Gram matrix. It is interesting to note that for BRIDGE, I could avoid assuming anything about the eigenvalues of the full matrix. Finally, as in the previous cases, the instruments are bounded in probability. 

With this set of assumptions, I prove the following two results. It is noteworthy that the objective in the first theorem is slightly stronger than model selection consistency. Instead of only identifying the true zeros, the method also identifies the sign of the relevant covariates. The second theorem is the standard oracle result from the previous two cases.

\begin{theorem}
\label{thm:large_msc_ad}
 Under (D.1-D.6), $P(\hat{\beta}_n=_s\beta_0)\to 1$. 
\end{theorem}

\begin{theorem}
\label{thm:large_oracle_ad}
     Under (D.1-D.6) and $0<\gamma<1$.  Let $s_{n}^2=\sigma_{\varepsilon}^2\delta_n'(\Sigma_{1n}^{d})^{-1}\delta_n$ for all vectors $\delta_n $ of size $(k_{xn}\times1)$ st. $\norm{\delta_n}\leq1$. Then, $n^{1/2}s_{n}^{-1}\delta_n' (\hat{\beta}_{1n}-\beta_{10})=n^{-1/2}s_{n}^{-1}\sum_{i=1}^{n}\varepsilon_{i}\delta_n'(\Sigma_{1n}^{d})^{-1}d_{1i}+o_p(1)\xrightarrow{d}N(0,1)$.

\end{theorem}


\mysection{3}{Computation}

\subsubsection*{Construction of Standard Errors}

In the two step problem, the researcher should first pick a tuning parameter $\gamma$ for each stage. In this exposition, in order to simplify the notation, I assumed the same $\gamma$ in both cases throughout the theoretical results. This comes without loss of generality analytically, but the implications of different choices will be exposed in a later subsection. After computing $\hat{\alpha}_n$ with BRIDGE, she should construct $\hat{D}_n$ and, plugging it in the 2nd stage, compute $\hat{\beta}_n$ with BRIDGE. Then, she can construct a consistent estimator of $s_{dn}$, which can be done in two parts: (1) estimate the Gram matrix of the estimated covariates with non-zero coefficients using $\hat{\alpha}_n$ which overlaps with $\hat{\Sigma}_{dn}$ as defined above, and (2) estimate $\sigma_{\varepsilon}^2$ using the residuals from the 2nd stage. The, $\hat{s}_{dnj}^{-1}$ gives an approximate standard error for $\hat{\beta}_j$ for each $j$.

\subsubsection*{Algorithm}

Given that BRIDGE is often challenging to compute as a non-convex problem, I provide a short discussion on the algorithm I use for the simulations, proposed by \cite{Huang:2010}:

In the simple case of the linear model, having only one-step estimators, we define the LS loss function as \(Q_n(\beta)=\frac{1}{2}\sum_{i=1}^{n}(Y_i-X_i\beta)^2,\) and the bridge penalized function as \(L_n(\beta)= Q_n(\beta)+\lambda \sum_{j=1}^{p}|\beta_j|^{\gamma},\ for\ 0<\gamma<1.\) I will follow the algorithm proposed by \citeN{Huang:2010}, as an improvement to the algorithm of \citeN{Huang:2008}. The former is more efficient as it does not require any approximation. To achieve this, the authors use the fact that the maximizer $\hat{\beta}_n$ of the following function: \[S_n(\beta,\theta)=Q_n(\beta)+\sum_{j=1}^{p}\theta_j^{1-\frac{1}{\gamma}}|\beta_j|+\tau_n\sum_{j=1}^{p}\theta_j\] is equal to the optimizer of  $L_n(\beta)$, and vice versa, under the constraint that $\hat{\theta}_j\geq 0,\ for\ j=1,\dots, p$ and the tuning parameter $\lambda= \tau_n^{1-\gamma}\gamma^{\gamma}(1-\gamma)^{\gamma-1},\ where\ \tau_n$ is the penalty parameter of $S_n(\beta,\theta)$ \cite{Huang:2009}.

Based on this result, they propose a simple iterative algorithm: First, I pick the initial value for $\beta^{(0)}$ to be the the corresponding LASSO estimate of the problem. Then, I compute the $\theta_j^{(s)}$ that appear in $S_n(\beta,\theta)$ as: \[\theta_j^{(s)}=\left(\frac{1-\gamma}{\tau_n\gamma}\right)^{\gamma}|\beta_j^{s-1}|^{\gamma},\ j=1,\dots,p.\]Using that value, I minimize \[Q_n(\beta)+\sum_{j=1}^{p}(\theta_j^{(s)})^{1-\frac{1}{\gamma}}|\beta_j|\] over $\beta$ to compute the new value of the vector $\beta^{(s)}$. Then, I repeat the last two steps until we achieve convergence, which is always attainable as $S_n(\beta,\theta)$ decreases in each step. To set the penalty parameters $\tau_n$, I use the functional form of $\lambda$ that the equivalence result requires. To compute $\lambda$ itself, I use cross-validation, splitting the sample in 5 folds.

Applying the general algorithm to my case, I compute the 1st stage coefficients, then estimate $\hat{D}_n$, I plug it in the 2nd stage and estimate $\hat{\beta}_n$ using the same algorithm.

\subsubsection*{Simulated Data Generating Process }

\noindent

Regarding the simulated DGP, starting from the first-stage parameters, i.e., the matrix $\alpha$, I draw its columns $\alpha_j,\text{ for}\  j=1,\dots, p_x$ as follows. For the column elements $\alpha_{jh}$, where  $h=1,\dots,p_{zn}$, I pick $\alpha_{jh} \in (-5,5)$ only for $k_{xn}$ entries and $\alpha_{jh}=0$ otherwise. I add a small normal random noise to all the entries. The sparsity structure of the 1st stage coefficient matrix has to satisfy that the number of overall relevant instruments ($\sum_{j=1}^{p_{zn}}k_{znj}$)  is at least as high as the number of the relevant covariates and, to avoid further identification issues, the same subset of instruments is only used for at most as many covariates as its cardinality. To keep the example clean, there is one relevant instrument per 2nd stage covariate with different degrees of relevance defined by the corresponding elements of $\alpha$. The 1st stage $\gamma$ for BRIDGE is set out to be $0.1$.

The instrumental variables are drawn from a multivariate normal, with a flexible correlation pattern among the relevant ones. That is,  $Z_i\sim\mathcal{N}_{p_{zn}}(0,\Sigma_{\bm{z}}).$ We define the variance-covariance matrix $\Sigma_{z}$ to have a Toeplitz structure, i.e. $\Sigma_z|_{jk}=\rho^{|j-k|},\ j,k\in[k_{xn}],\ \rho=0.7$ for the principal submatrix referring to the relevant instruments and the identity matrix for the rest:  \[\Sigma_{z} =\begin{pmatrix}
\Sigma_z^{k_{xn}\times k_{xn}} & 0 \\
0 & \bm{I}
\end{pmatrix}\]

Further, I pick a joint normal distribution for the i.i.d. error terms in the first and the second stage, $(\{v_{ij}\}_{j=1}^{p_x},u_i)$, and I choose the variance covaraince matrix to allow for non-trivial correlation terms between the two. That is, $(u_i, \{v_{ij}\}_{j=1}^{p_{xn}})\sim \mathcal{N}_{1+p_{xn}}(0,\Sigma_{uv})$, where \[\Sigma_{uv}=\begin{pmatrix}
\sigma_u & \sigma_{uv}^T \\
\sigma_{uv} & \sigma_v\bm{I}
\end{pmatrix}\] with $\sigma_u=\sigma_{\bm{v}}=\sqrt{0.5}$ and for $\sigma_{uv}= (\sigma_{uv^1},\dots, \sigma_{uv^{p_x}})$, some elements are equal to 0.4 and some 0.15.  These choices give us a positive-definite matrix. Lastly, The second stage vector of non-zero coefficients is also constructed from numbers in $(-5,5)$. Now, one can draw $X_{ij}$ as $X_{ij}=Z_i'\alpha_j+v_{ij}$ and the outcome variable as $Y_i=X_i'=\beta+u_i$.

\subsubsection*{Numerical Results for BRIDGE}
\noindent 

The first set of simulations presented below is using very small sample sizes to examine the elementary dynamics between the number of covariates and $n$, as well as the role of $\gamma$ in the problem. The second set is examining a larger sample size $n=1000$ with increasing number of covariates from $100$ to $1000$, as a more realistic scenario on where the method might be useful.

The first table has data from 200 simulations with information on the average root mean squared error (RMSE), the median RMSE and the number of variables that the method picks to be non-zero. I set $p_{xn}=p_{zn}=30$, $k_{xn}=k_{zn}=6$. I present results for OLS as a baseline, even though it is expected to work poorly in this environment, for LASSO, as it is a popular solution once the researcher faces a high dimensional environment and BRIDGE for three values of $\gamma$, to examine its significance. I pick three sample sizes to observe whether the method performs well on the ``difficult" case that the sample size is equal to the number of variables, on the case that the sample size is equal to the number of covariates plus the number of instruments, and on the case that the sample size is double this number and thus, an ``easier" case.

The setup described in the previous subsection is an environment where LASSO is model selection consistent and oracle efficient \cite{Gold:2020}. Even in this environment that does not take advantage of BRIDGE's validity under more flexible distributions, the two methods are fairly competitive. Looking at table \ref{tab:rmse_summary} below, one can see that OLS is performing very poorly even with slightly bigger sample size than the number of the covariates. Even though the mean MSE of LASSO and BRIDGE sharply drop once the sample size is slightly larger, OLS RMSE is marginally reduced. For a sample size of 30, the mean of the former two is significantly higher than the median, implying large outliers, but the two of them balance once I increase the sample size. LASSO picks, on average, a larger model for a small sample size, but once $n>30$, LASSO and BRIDGE have competitive values for RMSE and pick similar model sizes, close to the truth.

\begin{table}[ht]
  \centering
  \caption{Performance of methods for $p_{xn}=p_{zn}=30$, $k_{xn}=k_{znj}=6$}
  \label{tab:rmse_summary}
  \renewcommand{\arraystretch}{1.2}
  \begin{tabular}{@{} llccc @{}}
    \toprule
    \multirow{2}{*}{Sample Size} & \multirow{2}{*}{Method} &
    \multicolumn{3}{c}{Performance Metrics} \\
    \cmidrule(lr){3-5}
    & & Mean RMSE & Median RMSE & \#~Variables \\
    \midrule
    \multirow{5}{*}{\textbf{30}}
      & OLS          & 8.89 & 2.23 & 30 \\
      & LASSO        & 1.28 & 0.99 & 22.57 \\
      & BRIDGE ($\gamma=0.2$) & 1.80 & 0.74 & 8.94 \\
      & BRIDGE ($\gamma=0.5$) & 1.93 & 1.11 & 10.36 \\
      & BRIDGE ($\gamma=0.8$) & 1.44 & 0.61 & 9.03 \\
    \midrule
    \multirow{5}{*}{\textbf{60}}
      & OLS          & 5.92 & 1.73 & 30 \\
      & LASSO        & 0.03 & 0.02 & 6 \\
      & BRIDGE ($\gamma=0.2$) & 0.02 & 0.02 & 6 \\
      & BRIDGE ($\gamma=0.5$) & 0.12, &  0.02 & 8.62 \\
      & BRIDGE ($\gamma=0.8$) & 0.02 & 0.02  & 6 \\
    \midrule
    \multirow{5}{*}{\textbf{120}}
      & OLS          & 5.65 & 1.94 & 30 \\
      & LASSO        & 0.02 & 0.02 & 6 \\
      & BRIDGE ($\gamma=0.2$) & 0.01 & 0.01 & 6 \\
      & BRIDGE ($\gamma=0.5$) & 0.07 & 0.01 & 7.54 \\
      & BRIDGE ($\gamma=0.8$) & 0.01 & 0.01 & 6 \\
    \bottomrule
  \end{tabular}
\end{table}

Let $\hat{S}$ to be the set of indices of the 2nd stage covariates chosen by each method to be non-zero. Table \ref{tab:selection_probs} presents two values: the probability that  $\hat{S}$ contains the set of indices of the non-zero variables of the true model and the probability that the two quantities overlap. As expected, OLS always chooses a very big model, failing to estimate any of the true zeros. For $n=30$, LASSO picks larger models than it should and BRIDGE smaller, even though BRIDGE has a higher probability of picking the exact correct one. As the sample size grows, LASSO and BRIDGE always pick a set that contains the true one and, with a persistently very high probability, they pick exactly the true one. 

\begin{table}[ht]
  \centering
  \caption{Model–selection probabilities}
  \label{tab:selection_probs}
  \renewcommand{\arraystretch}{1.2}
  \begin{tabular}{@{} llcc @{}}
    \toprule
    \multirow{2}{*}{Sample Size} & \multirow{2}{*}{Method} &
    \multicolumn{2}{c}{Selection Performance} \\
    \cmidrule(lr){3-4}
    & & $\Pr(\text{True} \subseteq \hat{S})$ & $\Pr(\hat{S} = \text{True})$ \\
    \midrule
    \multirow{5}{*}{\textbf{30}}
      & OLS                & 1.00 & 0.00 \\
      & LASSO              & 1.00 & 0.12 \\
      & BRIDGE ($\gamma=0.2$) & 0.38 & 0.23 \\
      & BRIDGE ($\gamma=0.5$) & 0.38 & 0.20 \\
      & BRIDGE ($\gamma=0.8$) & 0.38 & 0.18 \\
    \midrule
    \multirow{5}{*}{\textbf{60}}
      & OLS                & 1.00 & 0.00 \\
      & LASSO              & 1.00 & 1.00 \\
      & BRIDGE ($\gamma=0.2$) & 1.00 & 1.00 \\
      & BRIDGE ($\gamma=0.5$) & 1.00 & 0.89 \\
      & BRIDGE ($\gamma=0.8$) & 1.00 & 1.00 \\
    \midrule
    \multirow{5}{*}{\textbf{120}}
      & OLS                & 1.00 & 0.00 \\
      & LASSO              & 1.00 & 1.00 \\
      & BRIDGE ($\gamma=0.2$) & 1.00 & 1.00 \\
      & BRIDGE ($\gamma=0.5$) & 1.00 & 0.94 \\
      & BRIDGE ($\gamma=0.8$) & 1.00 & 1.00 \\
    \bottomrule
  \end{tabular}
\end{table}

Further, it is interesting to examine further the role of $\gamma$ in the 2nd stage. In table \ref{tab:gamma-results}, I present results for $n=60$, $p_{xn}=30, p_{zn}=30$ and $k_{xn}=k_{znj}=6$ for 13 values of $\gamma$. For the first stage, the 1st stage $\gamma$ is set to be $0.1$ across all simulations. Note that as 2nd stage $\gamma$ grows, the penalty function approaches LASSO. It can be observed that even in the edge cases, close to $0$ and $1$, the median performance remains reasonably good, and the variables collected are the same with the true model. The only steep change is the increased mean RMSE for $\gamma$ around $0.6-0.75$, which may be incidental. There is no strong evidence that the choice of $\gamma$ significantly affects the performance of BRIDGE.

\begin{table}[htbp]
\centering
\caption{BRIDGE Two-Stage Results over 100 Simulations}
\label{tab:gamma-results}
\begin{tabular}{r r r r}
\toprule
\textbf{$\gamma$} & Mean RMSE & Median RMSE & \# Variables \\
\midrule
0.01 & 0.0203 & 0.0200 & 6.00 \\
0.10 & 0.0203 & 0.0202 & 6.00 \\
0.20 & 0.0203 & 0.0204 & 6.00 \\
0.25 & 0.0203 & 0.0204 & 6.00 \\
0.30 & 0.0204 & 0.0206 & 6.00 \\
0.40 & 0.0431 & 0.0203 & 6.00 \\
0.50 & 0.0769 & 0.0204 & 6.00 \\
0.60 & 0.1835 & 0.0212 & 6.00 \\
0.70 & 0.2721 & 0.0234 & 6.00 \\
0.75 & 0.2222 & 0.0240 & 6.00 \\
0.80 & 0.0218 & 0.0209 & 6.00 \\
0.90 & 0.0227 & 0.0226 & 6.00 \\
0.99 & 0.0247 & 0.0251 & 6.00 \\
\bottomrule
\end{tabular}
\end{table}

Regarding the set of simulations with a higher sample size ($n=1000$), there are data of up to 200 simulations, and I report the same summary statistics. I increase the number of the instruments to $100$ in every case, where the relevant ones remain 6. The number of covariates is either 100, 500, or 1000, creating approximately the same dynamics as the ``smaller $n$" case. 

By table \ref{tab:rmse_summary_large_bridge}, one can observe that OLS significantly under-performs, even in the first case, with $p_{xn}=10\%$ of $n$.Even if picking the correct $0$s was not the researcher's objective, the mean RMSE is much higher than both LASSO and BRIDGE. BRIDGE seems to have many outliers in this case, increasing the mean RMSE and the average number of selected variables but it may be incidental, since it does very well on the next two cases. LASSO does worse in picking the correct model, but is competitive on the average error it makes.

A compatible story can be said about table \ref{tab:selection_probs_large_bridge}. OLS and LASSO always pick a model which contains the true one, but almost never exactly the true one. BRIDGE on the other hand picks exactly the correct model almost always. In the first case, potential outliers that involve a much larger model than the correct drop the $P(\hat S= True)$ to 73\%, but for the other two cases it seems to be performing very well. It is important to iterate here that this is still an environment favorable to LASSO, since the errors are still following a normal distribution. 

Forthcoming sets of simulations will include errors that follow a sub-Gaussian but not normal distribution as well as PDFs with fatter tails, as well as the comparative results with adaptive LASSO. 

\begin{table}[ht]
  \centering
  \caption{Performance of methods for $p_{zn}=100$, $k_{xn}=6$}
  \label{tab:rmse_summary_large_bridge}
  \renewcommand{\arraystretch}{1.2}
  \begin{tabular}{@{} llccc @{}}
    \toprule
    \multirow{2}{*}{$p_{xn}$} & \multirow{2}{*}{Method} &
    \multicolumn{3}{c}{Performance Metrics} \\
    \cmidrule(lr){3-5}
    & & Mean RMSE & Median RMSE & \#~Variables \\
    \midrule
    \multirow{3}{*}{\textbf{100}}
      & OLS          & 4.34 & 1.41 & 100 \\
      & LASSO        & 0.031 & 0.031 & 25.51 \\
      & BRIDGE ($\gamma=0.5$) & 0.124 & 0.004 & 28.71 \\
    \midrule
    \multirow{3}{*}{\textbf{500}}
      & OLS          & 0.08 & 0.08 & 500 \\
      & LASSO        & 0.03 & 0.03 & 62.39 \\
      & BRIDGE ($\gamma=0.5$) & 0.002 &  0.002 & 6.00 \\
    \midrule
    \multirow{3}{*}{\textbf{1000}}
      & OLS          & 0.07 & 0.07 & 1000 \\
      & LASSO        & 0.03 & 0.03 & 72.67 \\
      & BRIDGE ($\gamma=0.5$) & 0.002 & 0.002 & 6.00 \\
    \bottomrule
  \end{tabular}
\end{table}

\begin{table}[ht]
  \centering
  \caption{Model–selection probabilities}
  \label{tab:selection_probs_large_bridge}
  \renewcommand{\arraystretch}{1.2}
  \begin{tabular}{@{} llcc @{}}
    \toprule
    \multirow{2}{*}{$p_{xn}$} & \multirow{2}{*}{Method} &
    \multicolumn{2}{c}{Selection Performance} \\
    \cmidrule(lr){3-4}
    & & $\Pr(\text{True} \subseteq \hat{S})$ & $\Pr(\hat{S} = \text{True})$ \\
    \midrule
    \multirow{3}{*}{\textbf{100}}
      & OLS                & 1.00 & 0.00 \\
      & LASSO              & 1.00 & 0.00 \\
      & BRIDGE ($\gamma=0.5$) & 1.00 & 0.73 \\
    \midrule
    \multirow{3}{*}{\textbf{500}}
      & OLS                & 1.00 & 0.00 \\
      & LASSO              & 1.00 & 0.00 \\
      & BRIDGE ($\gamma=0.5$) & 1.00 & 1.00 \\
    \midrule
    \multirow{3}{*}{\textbf{1000}}
      & OLS                & 1.00 & 0.00 \\
      & LASSO              & 1.00 & 0.00 \\
      & BRIDGE ($\gamma=0.5$) & 1.00 & 1.00 \\
    \bottomrule
  \end{tabular}
\end{table}

\FloatBarrier
\mysection{3}{Conclusion}

This paper is an exposition of how BRIDGE and adaptive LASSO can be used in a very popular environment, the linear model under the presence of endogeneity, when the researcher faces high dimensionality in both stages of the process. Facing a larger class of problems compared to the usual analysis, i.e. replacing the assumption of normal with sub-gaussian errors, I prove that both methods are model selection consistent and oracle efficient even when the number of covariates exceeds the sample size. For BRIDGE, I also prove that if the former is lower than the latter, the same properties hold without sub-gaussian errors. BRIDGE requires a slightly weaker set of assumptions to have the desirable properties, while adaptive LASSO is expected to be much faster computationally, so the methods are competitive on different fronts and the one that is recommended depends on the researcher's resources.


\clearpage
\bibliography{hd}
\vfill

\newpage

\begin{appendices}

\begin{lemma}\label{lem:lemma1}

For the 1st stage coefficients, it holds that, (1)$\norm{\alpha}\leq (p_{xn}\max_{j}k_{znj})^{1/2}a_1=O_p((p_{xn}\max_{j}k_{znj})^{1/2})$, (2) $\norm{\hat{\alpha}_n-\alpha}=O_p((p_{xn}/n)^{1/2})$, (3) $E[\norm{\hat{\alpha}_n-\alpha}^2]=O_p(\frac{p_{xn}\max_jk_{znj}}{n})$, (4) $E[\norm{\hat{\alpha}_n}^2]=O_p(p_{xn}\max_jk_{znj})$.

    \textbf{Proof}:

    This is a corollary of (A1)-(A4). Keeping the same notation, it holds that 

    $\norm{\hat{\alpha}_n-\alpha}\leq\norm{\hat{\alpha}_n-\alpha}_F=\sqrt{tr{(\hat{\alpha}_n-\alpha)(\hat{\alpha}_n-\alpha)'}}=\sqrt{\sum_{j=1}^{p_{zn}}\norm{\hat{\alpha}_{nj}-\alpha_j}^2}$.
    
    For each $j$, $s_{nj}^2=\sigma_{v_j}^2\delta_n'\Sigma_{1nj}^{-1}\delta_n\leq \sigma_{v_j}^2\tau_{1z}^{-1}$ by Rayleigh quotient, where $\norm{\delta_n}\leq 1$. Hence, $\delta_n'(\hat{\alpha}_{nj}-\alpha_j)=O_p(n^{-1/2}s_{nj})$ and $\sup_{\norm{\delta_n}\leq 1}\delta_n'(\hat{\alpha}_{nj}-\alpha_j)=O_p(n^{-1/2}\tau_{1z}^{-1/2})$. Thus, $\norm{\hat{\alpha}_n-\alpha}\leq O_p(n^{-1/2}\tau_{1z}^{-1/2}p_{xn}^{1/2})=O_p((p_{xn}/n)^{1/2})$, since $\tau_{1z}$ is a constant.

    Similarly, $\norm{\alpha}\leq \sqrt{\sum_{j=1}^{p_{xn}}\norm{\alpha_j}^2}\leq \sqrt{\sum_{j=1}^{p_{xn}}k_{znj}a_1^2}\leq \sqrt{p_{xn}\max_jk_{znj}}a_1=O_p((p_{xn}\max_{j}k_{znj})^{1/2})$.

    Also, by oracle efficiency (A4),

    $\sqrt{n}s_{nj}^{-1}\delta_n'(\hat{\alpha}_{nj}-\alpha_j)\xrightarrow{d}N(0,1)$ for every $j=1,\dots, p_{xn}$. So, \[AVar(\delta_n'(\hat{\alpha}_{nj}-\alpha_j))=\frac{s_{nj}^2}{n}+o_p(n^{-1})=
    \frac{\sigma_{v_j}^2}{n}\delta_n'\Sigma_{1nj}^{-1}\delta_n+o_p(n^{-1})\]

    Choose $\delta_n=e_h$ the h-th canonical vector for $h=1,\dots,k_{znj}$:
    \begin{align*}
        &AVar(e_h'(\hat{\alpha}_{nj}-\alpha_j))=V_{n,hh}\leq n^{-1}\sigma_{vj}^2(\Sigma_{1nj}^{-1})_{hh}+o_p(n^{-1})\ \forall h\\
        &tr(V_n)=\sum_{h=1}^{k_{znj}}(V_{n,hh})\leq n^{-1}\sigma_{vj}^2tr(\Sigma_{1nj}^{-1})+o_p(k_{znj}/n)\\
        &tr(\Sigma_{1nj}^{-1})=\sum_{h=1}^{k_{znj}}\lambda_{hnj}^{-1}\leq \frac{k_{znj}}{\tau_{1j}}\text{for $\lambda_{hnj}$ the h-th eigenvalue, st. }\lambda_{hnj}^{-1}\in\left(\frac{1}{\tau_{2j}},\frac{1}{\tau_{1j}}\right).
        \end{align*}

        Now, \begin{align*}
            E[\norm{\hat{\alpha}_{nj}-\alpha_j}^2]&=E[(\hat{\alpha}_{nj}-\alpha_j)'(\hat{\alpha}_{nj}-\alpha_j)]\\
            &= tr(E[(\hat{\alpha}_{nj}-\alpha_j)(\hat{\alpha}_{nj}-\alpha_j)'])=tr(AVar(\hat{\alpha}_{nj}-\alpha_j))+(E\norm{\hat{\alpha}_{nj}-\alpha_j})^2\\
            &\stackrel{(1)}{=}tr(AVar(\hat{\alpha}_{nj}-\alpha_j))+o_p(n^{-1})\leq n^{-1}\sigma_{vj}^2\frac{k_{znj}}{\tau_{1j}}+o_p(k_{znj}/n)
        \end{align*}

        where (1) holds by $\hat{\alpha}_n$ being asymptotically unbiased due to consistency and its asymptotic variance being bounded by bounded eigenvalues.
        
        Subsequently, 

        \begin{align*}
            &E[\norm{\hat{\alpha}_n-\alpha}^2]\leq E[\norm{\hat{\alpha}_n-\alpha}_F^2]= \sum_{j=1}^{p_{xn}}E[\norm{\hat{\alpha}_{nj}-\alpha_j}^2]\\
            &=\sum_{j=1}^{p_{xn}}n^{-1}\sigma_{vj}^2\frac{k_{znj}}{\tau_{1j}}+o_p(\frac{p_{xn}\max_jk_{znj}}{n})\\
            &\leq \frac{p_{xn}\max_j\sigma_{vj}\max_j k_{znj}}{n\tau_{1z}}+o_p(\frac{p_{xn}\max_jk_{znj}}{n})= O_p(\frac{p_{xn}\max_jk_{znj}}{n})=o(1)O_p(1)=o_p(1)
        \end{align*}
where the second to last equality holds by assumption (B3.c).

Also, 
\begin{align*}
    &\norm{\hat{\alpha}}^2=\norm{(\hat{\alpha}-\alpha)+\alpha}^2\leq \norm{\hat{\alpha}-\alpha}^2+2\norm{\hat{\alpha}-\alpha}\norm{\alpha}+\norm{\alpha}^2\\
    \Rightarrow & E(\norm{\hat{\alpha}}^2)\leq E[\norm{\hat{\alpha}-\alpha}^2]+2E[\norm{\hat{\alpha}-\alpha}\norm{\alpha}]+E[\norm{\alpha}^2]\\
    \Rightarrow & E(\norm{\hat{\alpha}}^2)\leq E[\norm{\hat{\alpha}-\alpha}^2]+2\norm{\alpha}E[\norm{\hat{\alpha}-\alpha}]+\norm{\alpha}^2\\
     \Rightarrow & E(\norm{\hat{\alpha}}^2)\leq o_p(1)+(a_1p_{xn}\max_jk_{znj})^{1/2}o_p(1)+a_1p_{xn}\max_jk_{znj}\\
     \Rightarrow & E(\norm{\hat{\alpha}}^2)\leq O_p(p_{xn}\max_jk_{znj})
\end{align*}
    
\end{lemma}

\begin{lemma}\label{lem:lemma2}
    Let $\kappa$ be a $p_{xn}\times1$ vector. Under (B1).
    \[E\sup_{\norm{u}\leq\delta}\left|\sum_{i=1}^{n}\varepsilon_id_i'\kappa\right|\leq \delta\sigma_{\varepsilon} n^{1/2} (\rho_{2}^z)^{1/2} \norm{\alpha}\]
    \[E\sup_{\norm{u}\leq\delta}\left|\sum_{i=1}^{n}\varepsilon_i\hat{d}_i'\kappa\right|\leq\delta\sigma_{\varepsilon} n^{1/2}(\rho_{2}^z)^{1/2}  E[\norm{\hat{\alpha}}^2]^{1/2}\]

    \textbf{Proof:} 

    Note that $\varepsilon_i=v_i'\beta_0+u_i$. So, $E[\varepsilon_i]=0$ and $Var[\varepsilon_i]=Var[u_i+ \sum_{j=1}^{p_{xn}}v_{ij}\beta_{0j}]= \sigma_u^2+\sum_{j=1}^{p_{xn}}\sum_{h=1}^{p_{xn}}\beta_{0j}\beta_{0h}\sigma_{v_jv_h}+2\sum_{j=1}^{p_{xn}}\beta_{0j}\sigma_{uv_j}=\sigma_{\varepsilon}^2$, constant over $i$.
    
    \begin{align*}
        E\sup_{\norm{\kappa}\leq\delta}\left|\sum_{i=1}^{n}\varepsilon_i\hat{d}_i'\kappa\right|^2 &\leq E\sup_{\norm{\kappa}\leq\delta}\norm{\kappa}\norm{\sum_{i=1}^{n}\varepsilon_i\hat{d}_i}^2 \\
        & \leq \delta^2 E\left[\sum_{i=1}^{n}\varepsilon_i\hat{d}_i'\sum_{i=1}^{n}\varepsilon_i\hat{d}_i\right]\\
        & = \delta^2 E\left[\varepsilon_1\hat{d}_1'\sum_{i=1}^{n}\varepsilon_i\hat{d}_i+\dots+\varepsilon_n \hat{d}_n'\sum_{i=1}^{n}\varepsilon_i\hat{d}_i\right]\\
        & \stackrel{(1)}{=} \delta^2 E\left[\varepsilon_1^2 \hat{d}_1'\hat{d}_1+\dots+\varepsilon_n^2 \hat{d}_n'\hat{d}_n\right]\\
        &=\delta^2\sigma_{\varepsilon}^{2}\sum_{i=1}^{n}E[\hat{d}_i'\hat{d}_i]=\delta^2\sigma_{\varepsilon}^2\sum_{i=1}^{n}E\left[\norm{\hat{d}_i'}^2\right]\\
        &=\delta^2\sigma_{\varepsilon}^2\sum_{i=1}^{n}E\left[\norm{Z_i'\hat{\alpha}}^2\right]\stackrel{(2)}{\leq} \delta^2\sigma_{\varepsilon}^2\sum_{i=1}^{n}\norm{Z_i'}^2E\left[\norm{\hat{\alpha}}^2\right]\\
        &=\delta^2\sigma_{\varepsilon}^2 n\rho_{2n}^z E\left[\norm{\hat{\alpha}}^2\right]=\delta^2\sigma_{\varepsilon}^2 n\rho_{2}^z E\left[\norm{\hat{\alpha}}^2\right]\\\
    \end{align*}
    where (1) holds by the exogeneity of the instruments ($\varepsilon _i\perp\hat{d}_i$) and  $E[\varepsilon_1\varepsilon_2]=0$ by (A1) and (B1), and (2) holds by $Z_i$ being fixed and Cauchy-Schwartz inequality.

    The result follows by Jensen's inequality.

    Note that for $d_i$ instead of $\hat{d}_i$, the result holds as:

    \[E\sup_{\norm{\kappa}\leq\delta}\left|\sum_{i=1}^{n}\varepsilon_id_i'\kappa\right|\leq \delta\sigma_{\varepsilon} n^{1/2} p_{zn}^{1/2} \norm{\alpha}\leq \delta\sigma_{\varepsilon} n^{1/2} (\rho_{2}^z)^{1/2} (p_{xn}\max_jk_{znj})^{1/2}a_1\] by $d_i$ being deterministic and Lemma (1).
\end{lemma}
\newpage

\textbf{Proof of Theorem \ref{thm:consistency}}

    By definition of $\hat{\beta}_n$ and for $\hat{d}_i=Z_i\hat{\alpha}$, where $\hat{\alpha}$ is an $(p_{zn}\times p_{xn}) $ estimated by BRIDGE in the first stage:

    \begin{align*}
        &\sum_{i=1}^{n}(Y_i-\hat{d}_i'\hat{\beta}_n)^2+\lambda_{xn}\sum_{j=1}^{p_{xn}}|\hat{\beta}_j|^{\gamma} \leq \sum_{i=1}^{n}(Y_i-\hat{d}_i'\beta_0)^2+\lambda_{xn}\sum_{j=1}^{p_{xn}}|\beta_0|^{\gamma}\\
        \Leftrightarrow & \sum_{i=1}^{n}(Y_i-\hat{d}_i'\hat{\beta}_n)^2\leq \sum_{i=1}^{n}(Y_i-\hat{d}_i'\beta_0)^2+\underbrace{\lambda_{xn}\sum_{j=1}^{p_{xn}}|\beta_0|^{\gamma}}_{\eta_n} \\
         \Leftrightarrow &  \sum_{i=1}^{n}(Y_i-\hat{d}_i'\hat{\beta}_n)^2 - \sum_{i=1}^{n}(Y_i-\hat{d}_i'\beta_0)^2 \leq \eta_n\\
         \Leftrightarrow & \sum_{i=1}^{n}[\hat{d}_i'(\hat{\beta}_n-\beta_0)]^2 +2\sum_{i=1}^{n}\varepsilon_i\hat{d}_i'(\beta_0-\hat{\beta}_n)\leq \eta_n
    \end{align*}

Let $\delta_n=n^{1/2}\left(\hat{\Sigma}_{dn}\right)^{1/2}(\hat{\beta}_n-\beta_0)$ and $\Delta_n=n^{-1/2}\left(\hat{\Sigma}_{dn}\right)^{-1/2}\hat{D}_n'$ and $\varepsilon_n$ the vector of $\varepsilon_i$.

Then, 
\begin{align*}
    &\sum_{i=1}^{n}[\hat{d}_i'(\hat{\beta}_n-\beta_0)]^2 +2\sum_{i=1}^{n}\varepsilon_i\hat{d}_i'(\beta_0-\hat{\beta}_n)=\\
    &=\left[n^{1/2}\hat{\Sigma}_{dn}^{1/2}(\hat{\beta}_n-\beta_0)\right]'\left[n^{1/2}\hat{\Sigma}_{dn}^{1/2}(\hat{\beta}_n-\beta_0)\right]+2\sum_{i=1}^{n}\varepsilon_i\hat{d}_i'(\beta_0-\hat{\beta}_n)\\
    &=\delta_n'\delta_n-2 \sum_{i=1}^{n}\varepsilon_i \hat{d}_i'\hat{\Sigma}_{dn}^{-1/2}n^{-1/2}n^{1/2}\hat{\Sigma}_{dn}^{1/2}(\hat{\beta}_n-\beta_0)\\
    & =\delta_n'\delta_n-2\left(n^{-1/2}\hat{\Sigma}_{dn}^{-1/2}\hat{D}_n'\varepsilon_n\right)'\delta_n\\
    & =\delta_n'\delta_n-2\left(\Delta_n\varepsilon_n\right)'\delta_n\leq \eta_n\\
    &\Leftrightarrow \delta_n'\delta_n-2\left(\Delta_n\varepsilon_n\right)'\delta_n \pm (\Delta_n\epsilon_n)(\Delta_n\epsilon_n)'  \leq \eta_n\\
    &\Leftrightarrow \norm{\delta_n-\Delta_n\varepsilon_n}^2-\norm{\Delta_n\varepsilon_n}^2\leq \eta_n\\
    &\Leftrightarrow \norm{\delta_n-\Delta_n\varepsilon_n}\leq\norm{\Delta_n\varepsilon_n} + \eta_n^{1/2}\\
    & \Leftrightarrow \norm{\delta_n}\leq \norm{\delta_n-\Delta_n\varepsilon_n}+\norm{\Delta_n\varepsilon_n}\leq 2\norm{\Delta_n\varepsilon_n} + \eta_n^{1/2}\\
    &\Rightarrow \norm{\delta_n}^2\leq 4\norm{\Delta_n\varepsilon_n} + \eta_n +2\norm{\Delta_n\varepsilon_n}\eta_n^{1/2}\\
    &\Rightarrow \norm{\delta_n}^2\leq 6\norm{\Delta_n\varepsilon_n} + 3\eta_n \\
\end{align*}
 Let $\Delta_i$ be the $i$-th row of $\Delta_n$. Then, $\Delta_n\varepsilon_n=\sum_{i=1}^{n}\varepsilon_i\Delta_i$. Also, by the i.i.d. assumption, $E[\varepsilon_i\varepsilon_j]=0$ for $i\neq j$. Therefore,

 \[E[\norm{\Delta_n\varepsilon_n}^2]=\sum_{i=1}^{n}E[\norm{\Delta_i}^2]E[\varepsilon_i^2]=\sigma_{\varepsilon}^2\sum_{i=1}^{n}E[\norm{\Delta_i}^2]=\sigma_{\varepsilon}^2 tr(I_{p_{xn}\times p_{xn}})=\sigma_{\varepsilon}^2p_{xn} \] as
 \[\Delta_n\Delta_n'= n^{-1/2}\left(\hat{\Sigma}_{dn}\right)^{-1/2}\hat{D}_n'\hat{D}_n\left(\hat{\Sigma}_{dn}\right)^{-1/2}n^{-1/2}= (\hat{D}_n'\hat{D}_n)^{-1/2}\hat{D}_n'\hat{D}_n(\hat{D}_n'\hat{D}_n)^{-1/2} =I_{p_{xn}\times p_{xn}}\]

    Thus, we have $ E[\norm{\delta_n}]^2\leq 6\sigma_{\varepsilon}^2p_{xn} + 3\eta_n$ and because $\eta_n=\lambda_{xn}\sum_{j=1}^{p_{xn}}|\beta_0|^{\gamma}$ is the number of $\beta_{0j}\neq 0$, so the rate that $\eta_n$ is $O(\lambda_{xn}k_{xn})$. I can write:

    \begin{align*}
        E[\delta_n'\delta_n]&=nE[(\hat{\beta}_n-\beta_0)'\hat{\Sigma}_{dn}(\hat{\beta}_n-\beta_0)]\leq 6\sigma_{\varepsilon}^2p_{xn} + O(\lambda_{xn}k_{xn})\\
        &\Leftrightarrow nE[(\hat{\beta}_n-\beta_0)'(\hat{\Sigma}_{dn}\pm \Sigma_{dn})(\hat{\beta}_n-\beta_0)]\leq 6\sigma_{\varepsilon}^2p_{xn} + O(\lambda_{xn}k_{xn})\\
        &\Leftrightarrow nE[(\hat{\beta}_n-\beta_0)'(\Sigma_{dn})(\hat{\beta}_n-\beta_0)]+nE[(\hat{\beta}_n-\beta_0)'(\hat{\Sigma}_{dn} -\Sigma_{dn})(\hat{\beta}_n-\beta_0)]\leq 6\sigma_{\varepsilon}^2p_{xn} + O(\lambda_{xn}k_{xn})\\
    \end{align*}

    where \begin{align*}
        \Sigma_{dn} -\hat{\Sigma}_{dn}&=n^{-1}D_n'D_n-n^{-1}\hat{D}_n'\hat{D}_n=n^{-1}[(Z'\alpha)'(Z'\alpha)-(Z'\hat{\alpha}_n)'(Z'\hat{\alpha}_n)]\\
        &\leq n^{-1}\norm{Z}^2|\alpha'\alpha-\hat{\alpha}'\hat{\alpha}|\leq \rho_{2n}^z\left|\norm{\alpha}^2-\norm{\hat{\alpha}_n}^2\right|\\
        &\leq  \rho_{2n}^z\left|\norm{\alpha}-\norm{\hat{\alpha}_n}\right|(\norm{\alpha}+\norm{\hat{\alpha}_n})\leq \rho_{2n}^z(\norm{\alpha}+\norm{\hat{\alpha}_n})\norm{\hat{\alpha}_n-\alpha}\\
        &\stackrel{(1)}{\leq} \rho_{2n}^z(2\norm{\alpha}+o_p(1))O_p((p_{xn}/n)^{1/2})\\
        &\leq \rho_{2n}^z2\norm{\alpha}O_p((p_{xn}/n)^{1/2})\\
        &\stackrel{(2)}{\leq} 2\rho_{2n}^z(p_{xn}\max_jk_{znj})^{1/2}a_1(p_{xn}/n)^{1/2}O_p(1)\\
        & = 2\rho_{2}^z p_{xn}a_1(\max_jk_{znj}/n)^{1/2}O_p(1)\stackrel{(3)}{=}o_p(1)O_p(1)=o_p(1)
    \end{align*}

    where (1) follows by Lemma (1), (2) holds by Lemma 2, (3) holds by assumption (B.3c). Then, $\norm{\Sigma_{dn} -\hat{\Sigma}_{dn}}=o_p(1)$ by CMT, so $\hat{\Sigma}_{dn}\xrightarrow{p}\Sigma_{dn}$.

    For the second term in LHS of $E[\delta_n'\delta_n]$, I have:
    \begin{align*}
        |(\hat{\beta}_n-\beta_0)'(\hat{\Sigma}_{dn} -\Sigma_{dn})(\hat{\beta}_n-\beta_0)|&=\norm{(\hat{\beta}_n-\beta_0)'\Sigma_{dn}^{1/2}(\Sigma_{dn}^{-1/2}(\hat{\Sigma}_{dn} -\Sigma_{dn})\Sigma_{dn}^{-1/2})\Sigma_{dn}^{1/2}(\hat{\beta}_n-\beta_0)}\\
        &\leq \norm{\Sigma_{dn}^{-1/2}(\hat{\Sigma}_{dn} -\Sigma_{dn})\Sigma_{dn}^{-1/2}}(\hat{\beta}_n-\beta_0)'\Sigma_{dn}(\hat{\beta}_n-\beta_0)\\
        &\leq \norm{\Sigma_{dn}^{-1}}\norm{\hat{\Sigma}_{dn} -\Sigma_{dn}}(\hat{\beta}_n-\beta_0)'\Sigma_{dn}(\hat{\beta}_n-\beta_0)\\
        &\leq (\rho_{1}^d)^{-1} p_{xn}(\max_jk_{znj}/n)^{1/2}O_p(1) (\hat{\beta}_n-\beta_0)'\Sigma_{dn}(\hat{\beta}_n-\beta_0) \\
        &= p_{xn}(\max_jk_{znj}/n)^{1/2}O_p(1) (\hat{\beta}_n-\beta_0)'\Sigma_{dn}(\hat{\beta}_n-\beta_0)\\
        &= p_{xn}(\max_jk_{znj}/n)^{1/2} (\hat{\beta}_n-\beta_0)'\Sigma_{dn}(\hat{\beta}_n-\beta_0)\\
        &=  o_p(1)(\hat{\beta}_n-\beta_0)'\Sigma_{dn}(\hat{\beta}_n-\beta_0)
    \end{align*}

due to the uniform bound of the eigenvalues of $\hat{\Sigma}_{dn} ,\Sigma_{dn}$.

    \begin{align*}
        &n(\hat{\beta}_n-\beta_0)'(\Sigma_{dn})(\hat{\beta}_n-\beta_0)+n(\hat{\beta}_n-\beta_0)'(\hat{\Sigma}_{dn} -\Sigma_{dn})(\hat{\beta}_n-\beta_0)  \\
        \geq & n(\hat{\beta}_n-\beta_0)'(\Sigma_{dn})(\hat{\beta}_n-\beta_0)-|n(\hat{\beta}_n-\beta_0)'(\hat{\Sigma}_{dn} -\Sigma_{dn})(\hat{\beta}_n-\beta_0)|\\
        \geq & n(\hat{\beta}_n-\beta_0)'(\Sigma_{dn})(\hat{\beta}_n-\beta_0)-n p_{xn}(\max_jk_{znj}/n)^{1/2} (\hat{\beta}_n-\beta_0)'\Sigma_{dn}(\hat{\beta}_n-\beta_0)\\
        \geq & n(1-(p_{xn}(\max_jk_{znj}/n)^{1/2} )\rho_{1}^d\norm{\hat{\beta}_n-\beta_0}^2
    \end{align*}

    Since the LHS is equal to $E[\delta_n'\delta_n]$ I can use the corresponding bound:
    \begin{align*}
        &n[(\hat{\beta}_n-\beta_0)'\hat\Sigma_{dn}(\hat{\beta}_n-\beta_0)]\geq n(1-(p_{xn}(\max_jk_{znj}/n)^{1/2} )\rho_{1}^d\norm{\hat{\beta}_n-\beta_0}^2\\
        \Leftrightarrow\ 
        & nE[(\hat{\beta}_n-\beta_0)'\hat{\Sigma}_{dn}(\hat{\beta}_n-\beta_0)]\geq n(1-(p_{xn}(\max_jk_{znj}/n)^{1/2} )\rho_{1}^dE\left[\norm{\hat{\beta}_n-\beta_0}^2\right]\\
        \Leftrightarrow \ 
        & E\left[\norm{\hat{\beta}_n-\beta_0}^2\right]\leq O\left(\frac{\lambda_{xn}k_{xn}+p_{xn}}{n(1-(p_{xn}(\max_jk_{znj}/n)^{1/2} )}\right)=O\left(\frac{\lambda_{xn}k_{xn}+p_{xn}}{n(1+o_p(1))}\right)\\
        &E\left[\norm{\hat{\beta}_n-\beta_0}^2\right]\leq O_p\left(\frac{\lambda_{xn}k_{xn}+p_{xn}}{n}O_p(1)\right)=     O_p\left(\underbrace{\frac{\lambda_{xn}k_{xn}+p_{xn}}{n}}_{\zeta}\right)\end{align*}

    By Markov's inequality: \begin{align*}
       & P\left(\norm{\hat{\beta}_n-\beta_0}^2\geq M\zeta\right)\leq \frac{E\left[\norm{\hat{\beta}_n-\beta_0}^2\right]}{M\zeta}\leq \frac{O(\zeta)}{M\zeta}=O\left(\frac{1}{M}\right)\\
       \Leftrightarrow\ &P\left(\norm{\hat{\beta}_n-\beta_0}\geq \sqrt{M\zeta}\right)\leq O\left(\frac{1}{M}\right)\\
       \Leftrightarrow\ &\norm{\hat{\beta}_n-\beta_0}\leq O_p\left(\frac{\lambda_{xn}k_{xn}+p_{xn}}{n}\right)^{1/2}=o_p(1)
    \end{align*}
 where the last inequality holds by (B3.b)

    For the second part of the theorem, I will show that $\norm{\hat{\beta}_n-\beta_0}=O_p(\frac{(p_{xn}k_{xn}\max_jk_{znj})^{1/2}}{\sqrt{n}})$.

    Let $r_n=\frac{\sqrt{n}}{(p_{xn}k_{xn}\max_jk_{znj})^{1/2}}$. Following the proof of the theorem 3.2.5 in \citeN{Vaart:1997}, for each $n$ I partition the parameter space of $(\hat{\beta}_n-\beta_0)$ into ``shells", i.e. \[S_{jn}=\{\beta:2^{j-1}\leq r_n\norm{\beta-\beta_0}<2^j\},\text{ for }j\in\mathbb{Z}.\]

    If $r_n\norm{\beta-\beta_0}$ is larger than $2^M$ for a given $M$, then $\hat{\beta}_n$ is in one of the shells with $j\geq M$. By the definition of $\hat{\beta}_n$ as a minimizer of $L_n(b)$, for every $\epsilon>0$:
    
    \begin{align*}
        &P(r_n\norm{\hat{\beta}_n-\beta_0}>2^M)=\sum_{j>M}P(\hat{\beta}_n\in S_{jn})\\
        \leq & \sum_{\substack{j>M \\ 2^j<\epsilon r_n}} P\left(\inf_{\beta\in S_{jn}}L_n(\beta)-L_n(\beta_0)\leq 0\right) +P(r_n\norm{\hat{\beta}_n-\beta_0}\geq \epsilon)
    \end{align*}

    where the 2nd term converges to 0 by $\hat{\beta}_n$ being consistent by the 1st part of the theorem. Now, I can write:
    \begin{align*}
        &L_n(\beta)-L_n(\beta_0)=\sum_{i=1}^{n}(Y_i-\hat{d}_i'\beta)^2+\lambda_{xn}\sum_{j=1}^{k_{xn}}|\beta_{1j}|^{\gamma}+\lambda_{xn}\sum_{j=1}^{m_{xn}}|\beta_{2j}|^{\gamma}-\sum_{i=1}^{n}(Y_i-\hat{d}_i'\beta_0)^2-\lambda_{xn}\sum_{j=1}^{k_{xn}}|\beta_{01j}|^{\gamma}\\
        &\geq \sum_{i=1}^{n}(Y_i-\hat{d}_i'\beta)^2+\lambda_{xn}\sum_{j=1}^{k_{xn}}|\beta_{1j}|^{\gamma}-\sum_{i=1}^{n}(Y_i-\hat{d}_i'\beta_0)^2-\lambda_{xn}\sum_{j=1}^{k_{xn}}|\beta_{01j}|^{\gamma}\\
        &= \sum_{i=1}^{n} (d_i'\beta_0+\varepsilon_i-\hat{d}_i'\beta)^2-\sum_{i=1}^{n} (d_i'\beta_0+\varepsilon_i-\hat{d}_i'\beta_0)^2+\lambda_{xn}\sum_{j=1}^{k_{xn}}\{|\beta_{1j}|^{\gamma}-|\beta_{01j}|^{\gamma}\}\\
        &= \sum_{i=1}^{n} (d_i'\beta_0+\varepsilon_i-d_i'\beta+(d_i'-\hat{d}_i')\beta)^2-\sum_{i=1}^{n} (d_i'\beta_0+\varepsilon_i-d_i'\beta_0+(d_i'-\hat{d}_i')\beta_0)^2+\lambda_{xn}\sum_{j=1}^{k_{xn}}\{|\beta_{1j}|^{\gamma}-|\beta_{01j}|^{\gamma}\}\\
        &=\sum_{i=1}^{n} [(d_i'(\beta_0-\beta)+\varepsilon_i)+(d_i'-\hat{d}_i')\beta]^2-\sum_{i=1}^{n} (\varepsilon_i+(d_i'-\hat{d}_i')\beta_0)^2+\lambda_{xn}\sum_{j=1}^{k_{xn}}\{|\beta_{1j}|^{\gamma}-|\beta_{01j}|^{\gamma}\}\\
        &= \sum_{i=1}^{n} (d_i'(\beta_0-\beta)+\varepsilon_i)^2-\sum_{i=1}^{n}\varepsilon_i^2+\lambda_{xn}\sum_{j=1}^{k_{xn}}\{|\beta_{1j}|^{\gamma}-|\beta_{01j}|^{\gamma}\}\\
        &+\underbrace{\sum_{i=1}^{n}2(d_i-\hat{d}_i')[d_i'(\beta_0-\beta)+\varepsilon_i-\varepsilon_i\beta_0]+\sum_{i=1}^{n}\left[((d_i-\hat{d}_i)'\beta_0)^2-((d_i-\hat{d}_i)'\beta)^2\right]}_{I_4}\\
        &= \sum_{i=1}^{n}(d_i'(\beta_0-\beta))^2-2\sum_{i=1}^{n}\varepsilon_id_i'(\beta-\beta_0)+\lambda_{xn}\sum_{j=1}^{k_{xn}}\{|\beta_{1j}|^{\gamma}-|\beta_{01j}|^{\gamma}\}+I_4\\
        &=I_1+I_2+I_3+I_4
    \end{align*}

Considering a specific cell $S_{jn}$, \[I_1=\sum_{i=1}^{n}(d_i'\beta_0-\beta))^2=\norm{D'(\beta_0-\beta)}^2=\norm{D}^2\norm{\beta_0-\beta}\geq n\rho_{1n}^{d}2^{2(j-1)}r_n^{-2}\]

\[I_3 = \lambda_{xn}\sum_{j=1}^{k_{xn}}\{|\beta_{1j}|^{\gamma}-|\beta_{01j}|^{\gamma}\}= \lambda_{xn}\gamma \sum_{j=1}^{k_{xn}}|\beta_{01j}^*|^{\gamma-1}sgn(\beta_{01j})(\beta_{1j}-\beta_{01j})\] 

for some $\beta_{01j}$ between $\beta_{1j}$ and $\beta_{01j}$. Now, by (A4), I know that $\beta_j$s are finite and bounded away from 0, and I am considering $\beta$s such that $\norm{\beta-\beta_0}\leq \epsilon $, so there exists $c_1$, st $\max_j |\beta_{01j}^*|^{\gamma-1}\leq c_1$

Then,
\begin{align*}
    &|I_3|\leq c_1\lambda_{xn}\gamma\sum_{j=1}^{k_{xn}}|\beta_{1j}-\beta_{01j}|\leq c_1\lambda_{xn}\gamma\sqrt{k_{xn}}\norm{\beta_{1j}-\beta_{01j}}\leq c_1 \lambda_{xn}\gamma\sqrt{k_{xn}}\frac{2^j}{r_n}\\
    \Rightarrow&I_3\geq -c_1\lambda_{xn}\gamma\sqrt{k_{xn}}\frac{2^j}{r_n}
\end{align*}

Hence, in each shell $S_{jn}$: 
\begin{align*}
    & L_n(\beta)-L_n(\beta_0)\geq n\rho_{1n}^{d}2^{2(j-1)}r_n^{-2}-c_1\lambda_{xn}\gamma\sqrt{k_{xn}}\frac{2^j}{r_n} -|I_2|-|I_4|\\
    \Rightarrow&  \inf_{\beta\in S_{jn}} L_n(\beta)-L_n(\beta_0)\geq n\rho_{1n}^{d}2^{2(j-1)}r_n^{-2}-c_1\lambda_{xn}\gamma\sqrt{k_{xn}}\frac{2^j}{r_n} -\sup_{\beta\in S_{jn}}|I_2|-\sup_{\beta\in S_{jn}}|I_4|\\
    \Rightarrow & P\left(\inf_{\beta\in S_{jn}}  L_n(\beta)-L_n(\beta_0) \leq 0\right)= P\left( n\rho_{1n}^{d}2^{2(j-1)}r_n^{-2}-c_1\lambda_{xn}\gamma\sqrt{k_{xn}}\frac{2^j}{r_n} \leq\sup_{\beta\in S_{jn}}|I_2|+\sup_{\beta\in S_{jn}}|I_4|\right)\\
    & \leq \frac{E[\sup_{\beta\in S_{jn}}|I_2|]+E[\sup_{\beta\in S_{jn}}|I_4|]}{n\rho_{1n}^{d}2^{2(j-1)}r_n^{-2}-c_1\lambda_{xn}\gamma\sqrt{k_{xn}}2^jr_n^{-1}}
\end{align*}

where the first inequality holds by Markov's inequality. For the first term in the numerator, I can use Lemma 2:

\begin{align*}
    &E[\sup_{\beta\in S_{jn}}|I_2|]= E\sup_{\beta\in S_{jn}}\left|\sum_{i=1}^{n}\varepsilon_id_i'(\beta-\beta_0)\right|\leq 2^{j+1}r_n^{-1}\sigma_{\varepsilon}n^{1/2}(\rho_{2n}^z)^{1/2}\norm{\alpha}\\
    \leq & 2^{j+1}r_n^{-1}\sigma_{\varepsilon}a_1(\rho_{2n}^z)^{1/2}(np_{xn}\max_jk_{znj} )^{1/2}=  2^{j+1}r_n^{-1}c_2(np_{xn}\max_jk_{znj} )^{1/2}
\end{align*}

\begin{align*}
    E[\sup_{\beta\in S_{jn}}|I_4|]&= E\sup_{\beta\in S_{jn} }\left|\sum_{i=1}^{n}2(d_i-\hat{d}_i')[d_i'(\beta_0-\beta)+\varepsilon_i-\varepsilon_i\beta_0]+\sum_{i=1}^{n}\left[((d_i-\hat{d}_i)'\beta_0)^2-((d_i-\hat{d}_i)'\beta)^2\right]\right|\\
    &\leq E\sup_{\beta\in S_{jn} }\left| \left[\sum_{i=1}^{n}2(d_i-\hat{d}_i')[d_i'(\beta_0-\beta)+\varepsilon_i-\varepsilon_i\beta_0]\right|+\sum_{i=1}^{n}\left|((d_i-\hat{d}_i)'\beta_0)^2\right|+\sum_{i=1}^{n}\left|((d_i-\hat{d}_i)'\beta)^2\right|\right]\\
    &\leq E\sup_{\beta\in S_{jn} } \left|\sum_{i=1}^{n}2(d_i-\hat{d}_i)'[d_i'(\beta_0-\beta)+\varepsilon_i-\varepsilon_i\beta_0]\right|\\
    &+E\sup_{\beta\in S_{jn} }\left[\sum_{i=1}^{n}\left|((d_i-\hat{d}_i)'\beta_0)^2\right|+\sum_{i=1}^{n}\left|((d_i-\hat{d}_i)'\beta)^2\right|\right]\\
    & \leq  2^{j+1}r_n^{-1}\sqrt{n}\sqrt{p_{xn}\max{k_{znj}}}\sqrt{\rho_{2n}^{z}}[2^jr_n^{-1}+\sqrt{k_{xn}}b_1+\sigma_{\varepsilon}]O_p(1)+n   2^{2j}r_n^{-2} o_p(1)
    \end{align*}
where the last inequality holds by the following:

\begin{align*}
    &E\sup_{\beta\in S_{jn} } \left|\sum_{i=1}^{n}2(d_i-\hat{d}_i)'[d_i'(\beta_0-\beta)+\varepsilon_i-\varepsilon_i\beta_0]\right|\\
    \leq & E\sup_{\beta\in S_{jn} } \left|\sum_{i=1}^{n}2(d_i-\hat{d}_i)'[d_i'(\beta_0-\beta)]\beta\right|+E\sup_{\beta\in S_{jn} } \left|\sum_{i=1}^{n}2\varepsilon_i(d_i-\hat{d}_i)'(\beta-\beta_0)\right|\\ 
    =&2\left(E\sup_{\beta\in S_{jn} }|[D_n'(\beta_0-\beta)]'(D_n-\hat{D}_n)\beta|+ E\sup_{\beta\in S_{jn}} |\varepsilon_n'(D_n-\hat{D}_n)'(\beta-\beta_0)|\right)\\
    \leq & 2\left(E\sup_{\beta\in S_{jn} }\norm{D_n}\norm{\beta_0-\beta}\norm{D_n-\hat{D}_n}\norm{\beta}+ E\sup_{\beta\in S_{jn}} \norm{\varepsilon_n}\norm{D_n-\hat{D}_n}\norm{\beta-\beta_0}\right)\\
    \stackrel{(1)}{\leq} & 2( 2^jr_n^{-1}\norm{\beta}\norm{D_n}E\norm{D_n-\hat{D}_n}+2^jr_n^{-1}E\norm{\varepsilon_n}\norm{D_n-\hat{D}_n})\\
    \stackrel{(2)}{\leq} & 2^{j+1}r_n^{-1}(2^jr_n^{-1}+\sqrt{k_{xn}}b_1)\sqrt{n\rho_2^d}E\norm{Z}\norm{\hat{a}_n-\alpha}+2^jr_n^{-1}E\norm{\varepsilon_n}\norm{Z}\norm{\hat{\alpha}_n-\alpha}\\
    \stackrel{(3)}{\leq} & 2^{j+1}r_n^{-1}(2^jr_n^{-1}+\sqrt{k_{xn}}b_1)\sqrt{n\rho_2^d}\sqrt{n\rho_{2n}^z}E\norm{\hat{a}_n-\alpha}+2^jr_n^{-1}\sqrt{n\rho_{2n}^{z}}E\norm{\varepsilon_n}\norm{\hat{\alpha}_n-\alpha}\\
    \leq & 2^{j+1}r_n^{-1}(2^jr_n^{-1}+\sqrt{k_{xn}}b_1)\sqrt{n\rho_2^d}\sqrt{n\rho_{2n}^z}\sqrt{E\norm{\hat{a}_n-\alpha}^2}+2^jr_n^{-1}\sqrt{n\rho_{2n}^{z}}\sqrt{E\norm{\varepsilon_n}^2}\sqrt{E\norm{\hat{\alpha}_n-\alpha}^2}\\
    \leq &2^{j+1}r_n^{-1}n\sqrt{\rho_{2n}^{z}}\sqrt{E\norm{\hat{\alpha}_n-\alpha}^2}[2^jr_n^{-1}+\sqrt{k_{xn}}b_1+\sigma_{\varepsilon}]\\
    \stackrel{(4)}{\leq}&2^{j+1}r_n^{-1}\sqrt{n}\sqrt{p_{xn}\max{k_{znj}}}\sqrt{\rho_{2n}^{z}}[2^jr_n^{-1}+\sqrt{k_{xn}}b_1+\sigma_{\varepsilon}]O_p(1)= 2^{j+1}r_n^{-1}n\sqrt{\rho_{2n}^{z}}[2^jr_n^{-1}+\sqrt{k_{xn}}b_1+\sigma_{\varepsilon}]o_p(1)
\end{align*}

where (1) holds by $D_n$ being a product of deterministic terms and $\beta$ being in shell $S_{jn}$. (2) holds by assumption (B4) which implies  $\norm{\beta}\leq \norm{\beta-\beta_0}+\norm{\beta_0}\leq 2^jr_n^{-1}+k_{xn}b_1$ and by the spectral norm of $D_n$ being equal to $\sqrt{n}\rho_{2n}^d\leq \sqrt{n}\rho_2^d$. (3) holds by $Z$ being fixed similarly to (2), and (4) holds by Lemma (1).

\begin{align*}
    &E\sup_{\beta\in S_{jn} } \left[\sum_{i=1}^{n}((d_i-\hat{d}_i)'\beta_0)^2+\sum_{i=1}^{n}((d_i-\hat{d}_i)'\beta)^2\right] =E\sup_{\beta\in S_{jn} }\left[\norm{(D_n-\hat{D}_n)'\beta_0}^2-\norm{(D_n-\hat{D}_n)'\beta}^2\right]\\
    =& E\sup_{\beta\in S_{jn} }[(D_n-\hat{D}_n)'\beta_0-(D_n-\hat{D}_n)'\beta]'[(D_n-\hat{D}_n)'\beta_0+(D_n-\hat{D}_n)'\beta]\\
    =&  E\sup_{\beta\in S_{jn} }[(D_n-\hat{D}_n)'(\beta_0-\beta)]'[(D_n-\hat{D}_n)'(\beta_0+\beta)]\leq E\sup_{\beta\in S_{jn} } \norm{D_n-\hat{D}_n}^2[(\beta_0-\beta)'(\beta_0+\beta)]\\
   = &E \sup_{\beta\in S_{jn} } \norm{D_n-\hat{D}_n}^2[\norm{\beta_0}^2-\norm{\beta}^2] \leq  E\sup_{\beta\in S_{jn} } \norm{D_n-\hat{D}_n}^2\norm{\beta_0-\beta}^2\\
   \leq & E \norm{D_n-\hat{D}_n}^2 2^{2j}r_n^{-2}\leq E [\norm{Z}^2\norm{\hat{\alpha}_n-\alpha}^2 2^{2j}r_n^{-2}]= \norm{Z}^22^{2j}E[\norm{\hat{\alpha}_n-\alpha}^2 r_n^{-2}]\\
  \leq & n \rho_{2n}^z 2^{2j}r_n^{-2}E[\norm{\hat{\alpha}_n-\alpha}^2 ]\leq n \rho_{2n}^z 2^{2j}r_n^{-2} o_p(1)\leq n  2^{2j}r_n^{-2} o_p(1)\\
\end{align*}
where the last inequality holds by Lemma \ref{lem:lemma1}.

Back to the initial objective:

\begin{align*}
   & P\left(\inf_{\beta\in S_{jn}}  L_n(\beta)-L_n(\beta_0) \leq 0 \right) \leq \frac{E[\sup_{\beta\in S_{jn}}|I_2|]+E[\sup_{\beta\in S_{jn}}|I_4|]}{n\rho_{1n}^{d}2^{2(j-1)}r_n^{-2}-c_1\lambda_{xn}\gamma\sqrt{k_{xn}}2^jr_n^{-1}}\\
    &\leq \frac{2^{j+1}r_n^{-1}\sqrt{n}\sqrt{p_{xn}\max{k_{znj}}}\sqrt{\rho_{2n}^{z}}[2^jr_n^{-1}+\sqrt{k_{xn}}b_1+\sigma_{\varepsilon}]O_p(1) }{n\rho_{1n}^{d}2^{2(j-1)}r_n^{-2}-c_1\lambda_{xn}\gamma\sqrt{k_{xn}}2^jr_n^{-1}}\\
    & +\frac{2^{j+1}r_n^{-1}c_2(np_{xn}\max_jk_{znj} )^{1/2}+n  2^{2j}r_n^{-2} o_p(1)}{n\rho_{1n}^{d}2^{2(j-1)}r_n^{-2}-c_1\lambda_{xn}\gamma\sqrt{k_{xn}}2^jr_n^{-1}}
\end{align*}

For $r_n=\frac{\sqrt{n}}{(p_{xn}k_{xn}\max_jk_{znj})^{1/2}}$, I will look each term of the numerator separately:

Term 1:

\begin{align*}
    &\frac{2^{j+1}r_n^{-1}\sqrt{n}\sqrt{p_{xn}\max{k_{znj}}}\sqrt{\rho_{2}^{z}}[2^jr_n^{-1}+\sqrt{k_{xn}}b_1+\sigma_{\varepsilon}]O_p(1) }{n\rho_{1n}^{d}2^{2(j-1)}r_n^{-2}-c_1\lambda_{xn}\gamma\sqrt{k_{xn}}2^jr_n^{-1}}\\
    \leq & \frac{2^{j+1}r_n^{-1}\sqrt{n}\sqrt{p_{xn}\max{k_{znj}}}\sqrt{\rho_{2}^{z}}2^jr_n^{-1}O_p(1)}{n\rho_{1n}^{d}2^{2(j-1)}r_n^{-2}-c_1\lambda_{xn}\gamma\sqrt{k_{xn}}2^jr_n^{-1}}+\frac{2^{j+1}r_n^{-1}\sqrt{n}\sqrt{p_{xn}\max{k_{znj}}}\sqrt{\rho_{2}^{z}}\sqrt{k_{xn}}b_1O_p(1)}{n\rho_{1n}^{d}2^{2(j-1)}r_n^{-2}-c_1\lambda_{xn}\gamma\sqrt{k_{xn}}2^jr_n^{-1}}\\
    &+\frac{2^{j+1}r_n^{-1}\sqrt{n}\sqrt{p_{xn}\max{k_{znj}}}\sqrt{\rho_{2}^{z}}\sqrt{k_{xn}}b_1O_p(1)}{n\rho_{1n}^{d}2^{2(j-1)}r_n^{-2}-c_1\lambda_{xn}\gamma\sqrt{k_{xn}}2^jr_n^{-1}}\\
    \leq & \frac{2^{2j+1}r_n^{-2}\sqrt{np_{xn}\max{k_{znj}}}O_p(1)}{n\rho_{1n}^{d}2^{2(j-1)}r_n^{-2}-c_1\lambda_{xn}\gamma\sqrt{k_{xn}}2^jr_n^{-1}}+\frac{2^{j+1}r_n^{-1}\sqrt{np_{xn}k_{xn}\max{k_{znj}}}O_p(1)}{n\rho_{1n}^{d}2^{2(j-1)}r_n^{-2}-c_1\lambda_{xn}\gamma\sqrt{k_{xn}}2^jr_n^{-1}}\\
    &+\frac{2^{j+1}r_n^{-1}\sqrt{np_{xn}\max{k_{znj}}}O_p(1)}{n\rho_{1n}^{d}2^{2(j-1)}r_n^{-2}-c_1\lambda_{xn}\gamma\sqrt{k_{xn}}2^jr_n^{-1}}\\
    \stackrel{(1)}{\leq}& \frac{2^{2j+1}no_p(1)}{n\rho_{1}^{d}2^{2(j-1)}-c_1\lambda_{xn}\gamma\sqrt{k_{xn}}2^jr_n}+\frac{2^{j+1}\sqrt{np_{xn}k_{xn}\max{k_{znj}}}O_p(1)}{n\rho_{1}^{d}2^{2(j-1)}r_n^{-1}-c_1\lambda_{xn}\gamma\sqrt{k_{xn}}2^j}+\frac{2^{j+1}\sqrt{np_{xn}\max{k_{znj}}}O_p(1)}{n\rho_{1}^{d}2^{2(j-1)}r_n^{-1}-c_1\lambda_{xn}\gamma\sqrt{k_{xn}}2^j}\\
    =&\frac{2^{2j+1}o_p(1)}{\rho_{1}^{d}2^{2(j-1)}-c_1\gamma\frac{\lambda_{xn}\sqrt{k_{xn}}}{\sqrt{n}}2^j\frac{1}{p_{xn}k_{xn}\max_jk_{znj}}}+\frac{2^{j+1}O_p(1)}{\frac{\sqrt{n}}{\sqrt{p_{xn}k_{xn}\max{k_{znj}}}}\rho_{1}^{d}2^{2(j-1)}r_n^{-1}-c_1\gamma\frac{\lambda_{xn}\sqrt{k_{xn}}}{\sqrt{n}}2^j\frac{1}{\sqrt{p_{xn}k_{xn}\max{k_{znj}}}}}\\
    &+\frac{2^{j+1}O_p(1)}{\frac{\sqrt{n}}{\sqrt{p_{xn}\max{k_{znj}}}}\rho_{1}^{d}2^{2(j-1)}r_n^{-1}-c_1\gamma\frac{\lambda_{xn}\sqrt{k_{xn}}}{\sqrt{n}}2^j\frac{1}{\sqrt{p_{xn}\max{k_{znj}}}}}\\
    \stackrel{(2)}{\leq}&   \frac{2^{2j+1}o_p(1)}{\rho_{1}^{d}2^{2(j-1)}-o_p(1)}+\frac{2^{j+1}O_p(1)}{\rho_{1}^{d}2^{2(j-1)}+o_p(1)}+\frac{2^{j+1}O_p(1)}{\sqrt{k_{xn}}\rho_{1}^{d}2^{2(j-1)}+o_p(1)}\\
    \stackrel{(3)}{\leq}&   \frac{O_p(1)}{2^{j-3}+o_p(1)}+o_p(1) \stackrel{(4)}{\leq} \frac{O_p(1)}{2^{j-4}}+o_p(1)
\end{align*}

 where (1) holds by (B.3c) and (2) holds by (B.3a). (3) holds by $k_{xn}$ growing but even when it is constant, I can use a similar argument as in the first term, i.e. note that  for $n$ large enough, the denominator $2^{j-3}+o_p(1)\geq 2^{j-4}$ for all $j\geq4$. 

Term 2:

\begin{align*}
    &\frac{2^{j+1}r_n^{-1}c_2(np_{xn}\max_jk_{znj} )^{1/2}}{n\rho_{1n}^{d}2^{2(j-1)}r_n^{-2}-c_1\lambda_{xn}\gamma\sqrt{k_{xn}}2^jr_n^{-1}}\leq \frac{2c_2(np_{xn}\max_jk_{znj} )^{1/2}}{n\rho_{1}^{d}2^{j-2}r_n^{-1}-c_1\gamma\lambda_{xn}\sqrt{k_{xn}}}\\
   = & \frac{2c_2}{\rho_{1}^{d}2^{j-2}r_n^{-1}\frac{\sqrt{n}}{\sqrt{p_{xn}\max_jk_{znj}}}-\frac{c_1\gamma\lambda_{xn}\sqrt{k_{xn}}}{(np_{xn}k_{xn}\max_jk_{znj})^{1/2}}}= \frac{2c_2}{\rho_{1}^{d}2^{j-2}\sqrt{k_{xn}}+o_p(1)}=o_p(1)
\end{align*}

where the second to last equality holds by (B.3a) and the last one holds by $k_{xn}$ grows. If $k_{xn}$ is a constant, the term is $o_p(1)$ by the same argument as in term 1.

 Term 3: 
\begin{align*}
    &\frac{n  2^{2j}r_n^{-2}o_p(1)}{n\rho_{1n}^{d}2^{2(j-1)}r_n^{-2}-c_1\lambda_{xn}\gamma\sqrt{k_{xn}}2^jr_n^{-1}} \leq \frac{n  2^{2j}r_n^{-2}o_p(1)}{n\rho_{1}^{d}2^{2(j-1)}r_n^{-2}-c_1\lambda_{xn}\gamma\sqrt{k_{xn}}2^jr_n^{-1}}\\
    = & \frac{ 2^{2j}o_p(1)}{\rho_{1}^{d}2^{2(j-1)}-c_1\gamma2^j\lambda_{xn}\sqrt{k_{xn}}r_n n^{-1}} = \frac{ \ 2^{2j}o_p(1)}{\rho_{1}^{d}2^{2(j-1)}-c_1\gamma 2^j\frac{\lambda_{xn}\sqrt{k_{xn}}}{\sqrt{n}(p_{xn}k_{xn}\max_jk_{znj})^{1/2}}}= o_p(1)\frac{2^j}{2^{j+1}+o_p(1)}=o_p(1)
\end{align*}

where the last equality holds by (B.3a).

To conclude the proof, \[ \sum_{\substack{j>M \\ 2^j<\epsilon r_n}} P\left(\inf_{\beta\in S_{jn}}L_n(\beta)-L_n(\beta_0)\leq 0\right)\leq \sum_{\substack{j>M }}\left(\frac{O_p(1)}{2^{j-4}}+o_p(1)\right)\leq  O_p(1)\sum_{\substack{j>M }}\frac{1}{2^{j-5}}\leq O_p(1)2^{-(M-6)}\]

which converges to $0$ for all $M=M_n\to\infty$. This completes the proof. 
\newpage

\begin{lemma}\label{lem:lemma3}
    Suppose $0<\gamma<1$. Let $\hat{\beta}_n=(\hat{\beta}_{1n}',\hat{\beta}_{2n}')'$. Under (A.1-A.4) and (B.1-B.4), $\hat{\beta}_{2n}=0$ with probability converging to 1.
\end{lemma}

   By Theorem \ref{thm:consistency}, for $C$ large enough, $\hat{\beta}_n$ is in the ball $\{\beta: \norm{\beta-\beta_0}\leq h_nC\}$ with probability converging to 1, as the probability of $\hat{\beta}$ being in a shell $>C$ goes to 0, where $h_n$ is the rate obtained in the 2nd part of Theorem 2. 

    Let $\beta_{1n}=\beta_{01}+h_nu_1$ \& $\beta_{2n}=\beta_{02}+h_nu_2=h_nu_2$, st. \begin{align*}
        &\norm{u}_2^2=\norm{u_1}_2^2+\norm{u_2}_2^2\leq C^2 \text{ as}\\
        &\norm{\beta-\beta_0}_2^2\leq h_n^2C^2\Leftrightarrow \norm{h_nu}_2^2=\norm{h_nu_1}_2^2+\norm{h_nu_2}_2^2=\norm{\beta_{1n}-\beta_{01}}_2^2+\norm{\beta_{2n}-\beta_{02}}_2^2\leq h_n^2C^2
    \end{align*}

    Let $V_n(u_1,u_2)=L_n(\beta_{1n},\beta_{2n})-L_n(\beta_{01},0)=L_n(\beta_{01}+h_nu_1,h_nu_2)-L_n(\beta_{01},0)$. Suffices to show that, for any $u_1$, $u_2$ with $\norm{u}\leq C$, if $\norm{u_2}>0$ then $V_n(u_1,u_2)-V_n(u_1,0)>0$ with probability converging to 1, as:
   
        \[V_n(u_1,u_2)-V_n(u_1,0)= L_n(\beta_{01}+h_nu_1,h_nu_2)-L_n(\beta_{01}+h_nu_1,0)\]i.e. the loss function under a strictly positive $\norm{u_2}=0$ is higher than the loss function where $\norm{u_2}=0$, with probability converging to 1. Equivalently, this implies that $\beta_{2n}=0$ with probability converging to 1 in the specific partition of the parameter space.

        Define $\hat{d}_{1i}$ the estimator of $d_{1i}$, the corresponding row of $D_{1n}$ and $\hat{d}_{2i}$ correspondingly. Let $m_{xn}=p_{xn}-k_{xn}$

        \begin{align*}
            &L_n(\beta_{01}+h_nu_1,h_nu_2)-L_n(\beta_{01}+h_nu_1,0)\\
            =&\sum_{i=1}^{n}(Y_i-\hat{d}_{1i}'(\beta_{01}+h_nu_1)-\hat{d}_{2i}'h_nu_2)^2-\sum_{i=1}^{n}(Y_i-\hat{d}_{1i}'(\beta_{01}+h_nu_1))^2+\lambda_{xn}\sum_{j=1}^{m_{xn}}|h_nu_2|^{\gamma}\\
            =& -2\sum_{i=1}^{n}(Y_i-\hat{d}_{1i}'(\beta_{01}+h_nu_1))'(\hat{d}_{2i}'h_nu_2)+\lambda_{xn}\sum_{j=1}^{m_{xn}}|h_nu_2|^{\gamma}\\
            =& -2 \sum_{i=1}^{n}(\varepsilon_i+(d_{1i}-\hat{d}_{1i})'\beta_{01}-\hat{d}_{1i}'h_nu_1)'(\hat{d}_{2i}'h_nu_2)+\sum_{i=1}^{n}(\hat{d}_2'h_nu_2)^2+\lambda_{xn}\sum_{j=1}^{m_{xn}}|h_nu_2|^{\gamma}\\
            =& h_n^2\sum_{i=1}^{n}(\hat{d}_{2i}'u_2)^2+2h_n^2\sum_{i=1}^{n}(\hat{d}_{2i}'u_2)(\hat{d}_{1i}'u_1)-2h_n\sum_{i=1}^{n}\varepsilon_i\hat{d}_{2i}'u_2+\lambda_{xn}\sum_{j=1}^{m_{xn}}|h_nu_2|^{\gamma}-2\sum_{i=1}^{n}(d_{1i}-\hat{d}_{1i})'\beta_{01}\hat{d}_{2i}'h_nu_2\\
            =& \mathbb{I}_1+\mathbb{I}_2+\mathbb{I}_3+\mathbb{I}_4+\mathbb{I}_5
            \end{align*}

        \begin{align*}
            \mathbb{I}_1+\mathbb{I}_2 &=h_n^2\sum_{i=1}^{n}(\hat{d}_{2i}'u_2)^2+2h_n^2\sum_{i=1}^{n}(\hat{d}_{2i}'u_2)(\hat{d}_{1i}'u_1)\\
            &= h_n^2\sum_{i=1}^{n}(\hat{d}_{2i}'u_2)^2+  h_n^2\sum_{i=1}^{n}(\hat{d}_{2i}'u_2-\hat{d}_{1i}'u_1)-h_n^2\sum_{i=1}^{n}(\hat{d}_{2i}'u_2)^2-h_n^2\sum_{i=1}^{n}(\hat{d}_{1i}'u_1)^2\\
            &\geq -h_n^2\sum_{i=1}^{n}(\hat{d}_{1i}'u_1)^2\geq -h_n^2C^2[\rho_{2n}^{z}p_{xn}O_p(1)+n\rho_{2n}^{d}]\geq -h_n^2[p_{xn}O_p(1)+nC^2\rho_{2}^{d}]
        \end{align*}
        
            where 
            \begin{align*}
                &h_n^2\sum_{i=1}^{n}(\hat{d}_{1i}'u_1)^2=\norm{\hat{D}_{1n}u_1}^2\leq \norm{\hat{D}_{1n}}^2\norm{u_1}^2\leq \norm{\hat{D}_{n}}^2\norm{u_1}^2\\
                &\leq \norm{Z}^2\norm{\hat{\alpha}}C^2
            \end{align*}
where $\norm{\hat{D}_{1n}}\leq\norm{\hat{D}_{n}}$ as $\hat{D}_{1n}'\hat{D}_{1n}$ is a principal matrix of $\hat{D}_n'\hat{D}_n$.

    For $\mathbb{I}_3= -2h_n\sum_{i=1}^{n}\varepsilon_1\hat{d}_{2i}'u_2$, I can write:
    \begin{align*}
        &E\left|\sum_{i=1}^{n}\varepsilon_i\hat{d}_{2i}'u_2\right|\leq \left[E\left(\sum_{i=1}^{n}\varepsilon_i\hat{d}_{2i}'u_2\right)^2\right]^{1/2}=\left[E\norm{\varepsilon_i\hat{D}_2u_2}^2\right]^{1/2}\\
        \leq & \left[E\left(\norm{\varepsilon_i}\norm{\hat{D}_2}\norm{u_2}\right)^2\right]^{1/2}\leq \left[E\sup_{\norm{u}\leq C}\left(\norm{\varepsilon_i}\norm{\hat{D}_2}\norm{u_2}\right)^2\right]^{1/2}\\
        \leq & C\sigma_{\varepsilon}n^{1/2}(\rho_{2n}^z)^{1/2}(E\norm{\hat{\alpha}_n}^2)^{1/2}= C\sigma_{\varepsilon}n^{1/2}(\rho_{2n}^z)^{1/2}\sqrt{p_{xn}\max_jk_{znj}}O_p(1)
    \end{align*}

    Then, $\mathbb{I}_3= h_nn^{1/2}(\rho_{2n}^z)^{1/2}\sqrt{p_{xn}\max_jk_{znj}}O_p(1)$

    For $\mathbb{I}_4= \lambda_{xn}\sum_{j=1}^{m_{xn}}|h_nu_{2j}|^{\gamma}=\lambda_{xn}|h_n|^{\gamma}O(\norm{u_2})^{\gamma}$, 
    I can write:\[ \sum_{j=1}^{m_{xn}}|u_2|^{\gamma}=\norm{u_{2j}}_{\gamma}^{\gamma}\Rightarrow \norm{u_{2}}_{\gamma}\geq \norm{u_{2}}_{2}\Rightarrow\norm{u_{2}}_{\gamma}^2\geq \norm{u_{2}}_{2}^2>0\]

    Then, $\mathbb{I}_4=\lambda_{xn}h_n^{\gamma}O(\norm{u_2}^2)$, where the last term of the product is bounded and strictly positive.

    The rate $\mathbb{I}_4/\mathbb{I}_1+\mathbb{I}_2$ goes to infinity because: $\lambda_{xn}h_n^{\gamma-2}/[p_{xn}O_p(1)+nC^2\rho_{2}^{d}] $ is the same rate as $\lambda_{xn}h_n^{\gamma}/n$ as $n$ grows faster than $p_{xn}$ by (B3.e). Replacing for $h_n$, I have $\frac{\lambda_{xn}n^{-\gamma/2}}{(\sqrt{p_{xn}k_{xn}\max_j k_{znj}})^{2-\gamma}}$ which goes to infinity by assumption (B.3f). Then, $\norm{u_2}>0$, implies that $V_n(u)>0$ with probability converging to 1.

\newpage

\textbf{Proof of Theorem \ref{thm:oracle}}
 
   Part (i) follows from Lemma \ref{lem:lemma3}. For part (ii), under (B.1-B.2) $\hat{\beta}_{1n}$ is consistent by Theorem 1. By (B.4), each component fo $\hat{\beta}_{1n}$ is bounded away from $0$ for $n$ sufficiently large. Then, at $(\hat{\beta}_{1n},\hat{\beta}_{2n})$, the following holds:
    \begin{align*}
        &\frac{\partial L_n(b_1,b_2)}{\partial b_1}\left|_{(\hat{\beta}_{1n},\hat{\beta}_{2n})}\right.=0\\
        &-\sum_{i=1}^{n}(Y_i-\hat{d}_{1i}'\hat{\beta}_{1n}-\hat{d}_{2i}'\hat{\beta}_{2n})\hat{d}_{1i}+\lambda_{xn}\gamma\sum_{j=1}^{p_{xn}}sgn(\hat{\beta}_{1nj})|\hat{\beta}_{1nj}|^{\gamma-1}=0
    \end{align*}
    and define a $(k_{xn}\times 1)$ vector $\psi_n$, which j-th element is $|\hat{\beta}_{1nj}|^{\gamma-1}sgn(\hat{\beta}_{1nj})$.
    Since $\hat{\beta}_{02}=0$ and $\varepsilon_i=Y_i-d_{1i}'\beta_{01}$:

    \begin{align*}
        &-\sum_{i=1}^{n}(Y_i-\hat{d}_{1i}'\hat{\beta}_{1n}-\hat{d}_{2i}'\hat{\beta}_{2n})\hat{d}_{1i}+\lambda_{xn}\gamma\psi_n=0\\
        \Leftrightarrow &-\sum_{i=1}^{n}(\varepsilon_i+d_{1i}'\beta_{01}\pm \hat{d}_{1i}'\beta_{01}-\hat{d}_{1i}'\hat{\beta}_{1n}-\hat{d}_{2i}'\hat{\beta}_{2n})\hat{d}_{1i}+\lambda_{xn}\gamma\psi_n=0\\
        \Leftrightarrow &-\sum_{i=1}^{n}(\varepsilon_i- \hat{d}_{1i}'(\hat{\beta}_{1n}-\beta_{01})-(d_{1i}+\hat{d}_{1i})'\hat{\beta}_{1n}-\hat{d}_{2i}'\hat{\beta}_{2n})\hat{d}_{1i}+\lambda_{xn}\gamma\psi_n=0\\
        \Leftrightarrow & \sum_{i=1}^{n} \hat{d}_{1i}'(\hat{\beta}_{1n}-\beta_{01})\hat{d}_{1i}= \sum_{i=1}^{n}\varepsilon_i \hat{d}_{1i}+ \sum_{i=1}^{n}(d_{1i}+\hat{d}_{1i})'\beta_{01}\hat{d}_{1i}-\sum_{i=1}^{n}\hat{d}_{2i}'\hat{\beta}_{2n}\hat{d}_{1i}+\lambda_{xn}\gamma\psi_n\\
        \Leftrightarrow & \hat{\Sigma}_{1d}(\hat{\beta}_{1n}-\beta_{01})= \sum_{i=1}^{n}\varepsilon_i \hat{d}_{1i}+ \sum_{i=1}^{n}(d_{1i}+\hat{d}_{1i})'\beta_{01}\hat{d}_{1i}-\sum_{i=1}^{n}\hat{d}_{2i}'\hat{\beta}_{2n}\hat{d}_{1i}+\lambda_{xn}\gamma\psi_n\\
        \Leftrightarrow &n^{1/2}\delta_n'(\hat{\beta}_{1n}-\beta_{01})=n^{-1/2}\sum_{i=1}^{n}\varepsilon_i\delta_n'(\hat\Sigma_{1n}^{d})^{-1}\hat{d}_{1i}+n^{-1/2}\sum_{i=1}^{n}\delta_n'(\hat\Sigma_{1n}^{d})^{-1}(d_{1i}-\hat{d}_{1i})'\beta_{01}\hat{d}_{1i}\\
        &-n^{-1/2}\sum_{i=1}^{n}\delta_n'(\hat\Sigma_{1n}^{d})^{-1}\hat{d}_{2i}'\hat{\beta}_{2n}\hat{d}_{1i}+n^{-1/2}\delta_n'(\hat\Sigma_{1n}^{d})^{-1}\lambda_{xn}\gamma\psi_n
    \end{align*}

    where $\norm{\delta_n}\leq 1$. Since $P(\hat{\beta}_{2n}=0)\to 1\Rightarrow  P(n^{-1/2}\sum_{i=1}^{n}\delta_n'(\Sigma_{1n}^{d})^{-1}\hat{d}_{2i}'\hat{\beta}_{2n}\hat{d}_{1i}=0)\to 1\Rightarrow n^{-1/2}\sum_{i=1}^{n}\delta_n'(\Sigma_{1n}^{d})^{-1}\hat{d}_{2i}'\hat{\beta}_{2n}\hat{d}_{1i}\xrightarrow{p}0$.

Note that since $\hat{\Sigma}_{dn}\xrightarrow{d}\Sigma_{dn}$, we also have $\hat{\Sigma}_{1n}^d\xrightarrow{p}\Sigma_{1n}^d$ by CMT. Then, the eigenvalues of $\hat{\Sigma}_{1n}^d$ converge in probability to the eigenvalues of $\Sigma_{1n}^d$, by Weyl's perturbation theorem. I.e. for all eigenvalues $\hat{\tau}_{hn}^d, \tau_{hn}^d$ of the corresponding matrices, we have $\forall h=1,\dots, p_{xn},\ |\hat{\tau}_{hn}^d-\tau_{hn}^d|\leq \max_h|\hat{\tau}_{hn}^d- \tau_{hn}^d|\leq \norm{\hat{\Sigma}_{1n}^d-\Sigma_{1n}^d}=o_p(1)$ by (B3.c). Then, $\hat{\tau}_{1n}^d\xrightarrow{p} \tau_{1n}^d$ and $\hat{\tau}_{2n}^d\xrightarrow{p} \tau_{2n}^d$

    For $n^{-1/2}\sum_{i=1}^{n}\delta_n'(\Sigma_{1n}^{d})^{-1}\hat{d}_{2i}'\hat{\beta}_{2n}\hat{d}_{1i}$:
    \begin{align*}
        &n^{-1/2}\left|\delta_n'(\hat{\Sigma}_{1n}^{d})^{-1}\psi_n\right|\leq n^{-1/2}\norm{\delta_n'}\norm{(\hat{\Sigma}_{1n}^{d})^{-1}}\norm{\psi_n}\\
        \leq & n^{-1/2} \norm{(\hat{\Sigma}_{1n}^{d})^{-1}} \norm{\psi_n}\leq n^{-1/2} \norm{(\hat{\Sigma}_{1n}^{d})^{-1}} k_{xn}b_0^{\gamma-1}\\
        = & n^{-1/2} \norm{\hat{\Sigma}_{1n}^{d}}^{-1} k_{xn}b_0^{\gamma-1}\stackrel{(1)}{\leq}n^{-1/2}\tau_{1}^{-1}k_{xn}b_0^{\gamma-1}+n^{-1/2}O_p(1)= {\leq}n^{-1/2}\tau_{1}^{-1}k_{xn}b_0^{\gamma-1}+o_p(1)=o_p(1).
        \end{align*}

where (1) holds by the spectral norm being equal to the largest eigenvalue, so the inverse is smaller than the smallest eigenvalue inverted. Let $\hat{\tau}_{1n}^d$ be the smallest eigenvalue of $\hat\Sigma_{1n}^{d}$, and $\hat\Sigma_{1n}^{d}$ is consistent. The last inequality holds by B.3.

Also,\begin{align*}
    &n^{-1/2}\sum_{i=1}^{n}\delta_n'(\hat\Sigma_{1n}^{d})^{-1}(d_{1i}-\hat{d}_{1i})'\beta_{01}\hat{d}_{1i}\leq n^{-1/2}|\delta_n'(\hat\Sigma_{1n}^{d})^{-1}\hat{D}_{1n}(D_{1n}-\hat{D}_{1n})'\beta_{01}|\\
    =\ & n^{-1/2}\norm{\delta_n'(\hat\Sigma_{1n}^{d})^{-1}\hat{D}_{1n}(D_{1n}-\hat{D}_{1n})'\beta_{01}}\leq \norm{\delta_n}\norm{(\hat\Sigma_{1n}^{d})^{-1}}\norm{\hat{D}_{1n}}\norm{(D_{1n}-\hat{D}_{1n})}\norm{\beta_{01}}\\
    \leq\ &n^{-1/2}\tau_{1}^{-1}k_{xn}\sqrt{n\tau_2}\norm{D_{1n}-\hat{D}_{1n}}=\tau_{1}^{-1}\tau_2^{1/2}k_{xn}\norm{D_{1n}-\hat{D}_{1n}}
\end{align*}

For the quantity of interest, I also have to do some decomposition:
\begin{align*}
    &n^{-1/2}\sum_{i=1}^{n}\varepsilon_i\delta_n'(\hat\Sigma_{1n}^{d})^{-1}\hat{d}_{1i}=n^{-1/2}\sum_{i=1}^{n}\varepsilon_i\delta_n'((\hat\Sigma_{1n}^{d})^{-1}\pm(\Sigma_{1n}^{d})^{-1})\hat{d}_{1i}\\
    =& n^{-1/2}\sum_{i=1}^{n}\varepsilon_i\delta_n'(\Sigma_{1n}^{d})^{-1}\hat{d}_{1i}+n^{-1/2}\sum_{i=1}^{n}\varepsilon_i\delta_n'((\hat\Sigma_{1n}^{d})^{-1}-(\Sigma_{1n}^{d})^{-1})\hat{d}_{1i}\\
    =&n^{-1/2}\sum_{i=1}^{n}\varepsilon_i\delta_n'(\Sigma_{1n}^{d})^{-1}d_{1i}+\underbrace{n^{-1/2}\sum_{i=1}^{n}\varepsilon_i\delta_n'(\Sigma_{1n}^{d})^{-1}(\hat d_{1i}-d_{1i})+n^{-1/2}\sum_{i=1}^{n}\varepsilon_i\delta_n'((\hat\Sigma_{1n}^{d})^{-1}-(\Sigma_{1n}^{d})^{-1})\hat{d}_{1i}}_{show \ o_p(1)}\\
\end{align*}

I will look at both terms separately:

For the first one, define \[T_n=n^{-1/2}((\Sigma_{1}^d)^{-1}-(\hat\Sigma_{1}^d)^{-1})D_{1n}\varepsilon_n=n^{-1/2}\sum_{i=1}^{n}\varepsilon_i\delta_n'((\Sigma_{1}^d)^{-1}-(\hat\Sigma_{1}^d)^{-1})d_{1i}=n^{-1/2}\sum_{i=1}^{n}\varepsilon_iu_i\] where $\norm{\delta_n}=1$.
Note that $E(T_n|u_n)=0$ and $Var(T_n|u_n)=n^{-1}\sigma_{\varepsilon}^2\sum_{i=1}^{n}u_{i}^2$.
Also, \begin{align*}
    n^{-1}\sum_{i=1}^{n}u_{i}^2&=n^{-1}\sum_{i=1}^{n}(\delta_n'((\Sigma_{1}^d)^{-1}-(\hat\Sigma_{1}^d)^{-1})d_{1i})^2\\
    &= \delta_n'((\Sigma_{1}^d)^{-1}-(\hat\Sigma_{1}^d)^{-1})\left(\frac{1}{n}\sum_{i=1}^{n}d_{1i}'d_{1i}\right)\delta_n'((\Sigma_{1}^d)^{-1}-(\hat\Sigma_{1}^d)^{-1})\\
    &\leq \norm{(\Sigma_{1}^d)^{-1}-(\hat\Sigma_{1}^d)^{-1}}^2\norm{\frac{1}{n}\sum_{i=1}^{n}d_{1i}'d_{1i}}^2\leq \norm{(\Sigma_{1}^d)^{-1}}^2\norm{(\hat\Sigma_{1}^d)^{-1}}^2\norm{\Sigma_{1}^d-\hat\Sigma_{1}^d}^2\norm{\frac{1}{n}\sum_{i=1}^{n}d_{1i}'d_{1i}}^2\\
    &\leq (\tau_{1}^{d}+o_p(1))^{-2}({\tau}_{1}^{d})^{-2}\frac{k_{xn}\sqrt{\max_jk_{znj}}}{\sqrt{n}}O_p(1)\tau_{2d}^{2}=o_p(1)
\end{align*}

where the last inequality holds by the uniform bounds of the eigenvalues and the norm of the difference of the ``relevant" Gram matrices being an $o_p(1)$, shown by an argument similar to the one in \ref{thm:consistency} for the full Gram matrix and (B.3c).

For $ n^{-1/2}\sum_{i=1}^{n}\varepsilon_i \delta_n'(\Sigma_{1n}^{d})^{-1}(\hat d_{1i}- d_{1i})$, I can use Chebyshev's inequality as follows:

\[T_n=n^{-1/2}\sum_{i=1}^{n}\varepsilon_i \delta_n'(\Sigma_{1n}^{d})^{-1}(\hat d_{1i}- d_{1i})=n^{-1/2}\sum_{i=1}^{n}\varepsilon_iu_i\] where $u_i=\delta_n'(\Sigma_{1n}^{d})^{-1}(\hat d_{1i}- d_{1i})$ and $\hat{D}_1-D_1=[\Delta'Z_i]_1=[(\hat\alpha-\alpha)'Z_i]_1$,. I also have that $\norm{(\Sigma_{1n}^{d})^{-1}}\leq\tau_1^{-1}$ and $\norm{\hat\alpha-\alpha}\leq O_p(\sqrt\frac{k_{xn}\max_j k_{znj}}{n})=o_p(1)$.

Then, $E[T_n|u_n]=0$ because $E[\varepsilon_i|u_n]=0$ by the exogeneity of the instruments and i.i.d. errors in both stages  and \[Var[T_n|u_n]=\frac{1}{n}\sum_{i=1}^{n}u_i^2Var(\varepsilon_i|u_n)\leq \frac{\sigma_{\varepsilon}^2}{n}\sum_{i=1}^{n}u_i^2\]

\[\forall \epsilon>0,\ P(|T_n|>\epsilon|u_n)\leq \frac{Var(T_n|u_n)}{\epsilon^2}\leq \frac{\sigma_{\varepsilon}^2}{n\epsilon^2}\sum_{i=1}^{n}u_i^2=\psi_n(u_n)\]
where
\begin{align*}
    \frac{1}{n}\sum_{i=1}^{n}u_i^2&= \frac{1}{n}\sum_{i=1}^{n}(\delta_n'(\Sigma_{1n}^{d})^{-1}(\hat d_{1i}- d_{1i}))^2=\delta_n'(\Sigma_{1n}^{d})^{-1}\Delta'\left(\frac{1}{n}\sum_{i=1}^{n}Z_i'Z_i\right)_1\Delta(\Sigma_{1n}^{d})^{-1}\delta_n\\
    & \leq \norm{(\Sigma_{1n}^{d})^{-1}} ^2\norm{\Delta}^2\norm{\frac{1}{n}\sum_{i=1}^{n}Z_i'Z_i}_1\leq  \tau_{2z}\tau_{1d}^{-2}\frac{k_{xn}\max_j k_{znj}}{n}O_p(1)=o_p(1)
\end{align*}
For every $\epsilon$, I can find $\delta$, such tha t:
\begin{align*}
    P(|T_n|>\epsilon)&= P(\{|T_n|>\epsilon\}\cap\{|\frac{1}{n}\sum_{i=1}^{n}u_i^2|>\mu\} )+P(\{|T_n|>\epsilon\}\cap\{|\frac{1}{n}\sum_{i=1}^{n}u_i^2|\leq\mu\}  )\\
    &\leq  P(|\frac{1}{n}\sum_{i=1}^{n}u_i^2|>\mu )+  \frac{\sigma_{\varepsilon}^2}{n\epsilon^2}\sum_{i=1}^{n}u_i^2\leq \frac{\delta}{2}+  \frac{\sigma_{\varepsilon}^2\mu}{\epsilon^2}\leq \delta
\end{align*}

Back to the quantity of interest:
\[n^{1/2}\delta_n'(\hat{\beta}_{1n}-\beta_{01})=n^{-1/2}\sum_{i=1}^{n}\varepsilon_i\delta_n'(\Sigma_{1n}^{d})^{-1}d_{1i}+o_p(1)\] or its standardized equivalent:
\[n^{1/2}s_{dn}^{-1}\delta_n'(\hat{\beta}_{1n}-\beta_{01})=n^{-1/2}s_{dn}^{-1}\sum_{i=1}^{n}\varepsilon_i\delta_n'(\Sigma_{1n}^{d})^{-1}d_{1i}+o_p(1)\]

To see why $s_n^2$ is the average variance over $n$, consider the following quantities: $v_i=n^{-1/2}s_{dn}^{-1}\delta_n'(\Sigma_{1n}^{d})^{-1}d_{1i}$ and $w_i=\varepsilon_iv_i$, so:
\begin{align*}
  Var\left(\sum_{i=1}^{n}\varepsilon_iv_i\right)& = Var \left(\sum_{i=1}^{n}\varepsilon_in^{-1/2}s_{dn}^{-1}\delta_n'(\Sigma_{1n}^{d})^{-1}d_{1i}\right)\\
  &=n^{-1}s_{dn}^{-2}\delta_n'(\Sigma_{1n}^{d})^{-1}Var\left(\sum_{i=1}^{n}\varepsilon_id_{1i}\right)(\Sigma_{1n}^{d})^{-1}\delta_n\\
  &= n^{-1}s_{dn}^{-2}\delta_n'(\Sigma_{1n}^{d})^{-1}\left(\sum_{i=1}^{n}d_{1i}Var\left(\varepsilon_i\right)d_{1i}'\right)(\Sigma_{1n}^{d})^{-1}\delta_n\\
  &= s_{dn}^{-2}\sigma_{\varepsilon}^2\delta_n'(\Sigma_{1n}^{d})^{-1}n^{-1}D_{1n}'D_{1n}(\Sigma_{1n}^{d})^{-1}\delta_n\\
  &=s_{dn}^{-2}\sigma_{\varepsilon}^2\delta_n'(\Sigma_{1n}^{d})^{-1}\delta_n=1
\end{align*}

To apply the Lindeberg-Feller CLT, I need to show that the Lindeberg condition holds for the RV $\varepsilon_i\delta_n'(\Sigma_{1n}^{d})^{-1}d_{1i}$, ie.:

\begin{align*}
    &\lim_{n\to \infty}\frac{1}{ns_{dn}^2}\sum_{i=1}^{n}E[(\varepsilon_i\delta_n'(\Sigma_{1n}^{d})^{-1}d_{1i})^2\mathbb{I}\{|\varepsilon_i\delta_n'(\Sigma_{1n}^{d})^{-1}d_{1i}|>\epsilon\sqrt{n}s_{dn}\}]=0\\
    \Leftrightarrow & \lim_{n\to \infty}\sum_{i=1}^{n}E[(n^{-1/2}s_{dn}^{-1}\varepsilon_i\delta_n'(\Sigma_{1n}^{d})^{-1}d_{1i})^2\mathbb{I}\{|n^{-1/2}s_{dn}^{-1}\varepsilon_i\delta_n'(\Sigma_{1n}^{d})^{-1}d_{1i}|>\epsilon\}]=0\\
    \Leftrightarrow & \lim_{n\to \infty}\sum_{i=1}^{n}E[w_i^2\mathbb{I}\{|w_i|>\epsilon\}]=0\Leftrightarrow \sigma_{\varepsilon}^2\lim_{n\to \infty}\sum_{i=1}^{n}E[w_i^2\mathbb{I}\{|w_i|>\epsilon\}]=0\\
     \Leftrightarrow & \lim_{n\to \infty}\sigma_{\varepsilon}^2\sum_{i=1}^{n}v_i^2E[\varepsilon_i^2\mathbb{I}\{|\varepsilon_iv_i|>\epsilon\}]
\end{align*}

Note that
\begin{align*}
&\sigma_{\varepsilon}^2\sum_{i=1}^{n}v_i^2=\sigma_{\varepsilon}^2\sum_{i=1}^{n}(n^{-1/2}s_{dn}^{-1}\delta_n'(\Sigma_{1n}^{d})^{-1}d_{1i})^2= n^{-1}s_{dn}^{-2}\sigma_{\varepsilon}^2\sum_{i=1}^{n}\delta_n'(\Sigma_{1n}^{d})^{-1}d_{1i}d_{1i}'(\Sigma_{1n}^{d})^{-1}\delta_n\\
&=s_{dn}^{-2}\sigma_{\varepsilon}^2\delta_n'(\Sigma_{1n}^{d})^{-1}(\Sigma_{1n}^{d})(\Sigma_{1n}^{d})^{-1}\delta_n=s_{dn}^{-2}s_{dn}^{2}=1
\end{align*}

Then, \begin{align*}
\sigma_{\varepsilon}^2\sum_{i=1}^{n}v_i^2E[\varepsilon_i^2\mathbb{I}\{|\varepsilon_iv_i|>\epsilon\}]&\leq \sigma_{\varepsilon}^2\sum_{i=1}^{n}v_i^2\max_{1\leq i\leq n}E[\varepsilon_i^2\mathbb{I}\{|\varepsilon_iv_i|>\epsilon\}]\\
&= \max_{1\leq i\leq n}E[\varepsilon_i^2\mathbb{I}\{|\varepsilon_iv_i|>\epsilon\}]
\end{align*}
so it suffices for the latter to go to $0$ as $n$ grows, for the condition to hold. 

Equivalently, I can write: 
\[\max_{1\leq i\leq n}|v_i|=n^{-1/2}s_{dn}^{-1}\max_{1\leq i\leq n}|\delta_n'(\Sigma_{1n}^{d})^{-1}d_{1i}|,\] where \[|\delta_n'(\Sigma_{1n}^{d})^{-1}d_{1i}|\leq (\delta_n'(\Sigma_{1n}^{d})^{-1}\delta_n)^{1/2}(d_{1i}'(\Sigma_{1n}^{d})^{-1}d_{1i})^{1/2}\] by the Cauchy-Schwartz and $(\Sigma_{1n}^{d})^{-1}$ being symmetric.

Also, $s_{dn}^{-1}=\sigma_{\varepsilon}^{-1}(\delta_n'(\Sigma_{1n}^{d})^{-1}\delta_n)^{-1/2}$. Then,

\begin{align*}
    \max_{1\leq i\leq n}|v_i|&\leq n^{-1/2}\sigma_{\varepsilon}^{-1}(\delta_n'(\Sigma_{1n}^{d})^{-1}\delta_n)^{-1/2}(\delta_n'(\Sigma_{1n}^{d})^{-1}\delta_n)^{1/2}\max_{1\leq i\leq n}(d_{1i}'(\Sigma_{1n}^{d})^{-1}d_{1i})^{1/2}\\
    &\leq n^{-1/2}\sigma_{\varepsilon}^{-1} \tau_{1}^{-1/2}\max_{1\leq i\leq n}(d_{1i}'d_{1i})^{-1/2}\to 0
\end{align*}
where the last relation holds by (B.5). Thus, Lindeberg condition holds and:
\[n^{-1/2}\sum_{i=1}^{n}\varepsilon_is_{dn}^{-1}\delta_n(\Sigma_{1n}^{d})^{-1}d_{1i}\xrightarrow{d}N(0,1)\] and, equivalently,
\[n^{1/2}s_{dn}^{-1}\delta_n'(\hat{\beta}_{1n}-\beta_{01})\xrightarrow{d}N(0,1).\]
\newpage
\begin{lemma}
  \label{lemma:KnightFu}
    \textbf{Knight and Fu(2000)}: Let $g(u)=u^2-2au+\lambda|u|^{\gamma}$, where $a\neq 0$, $\lambda\geq 0$, and $0<\gamma<1$. Denote \[c_{\gamma}=\left(\frac{2}{2-\gamma}\right)\left(\frac{2(1-\gamma)}{2-\gamma}\right)^{1-\gamma}.\] Then, $\argmin (g)=0$ iff $\lambda>c_{\gamma}|a|^{2-\gamma}$.
\end{lemma}

\textbf{Proof of Theorem \ref{thm:large_msc}}

 $\xi_{nj}=n^{-1}\sum_{i=1}^{n}(d_{1i}'\beta_{10})d_{ij}$. Let $a_j=(d_{1j},\dots,d_{nj})'$.

\begin{align*}
    U_n(\beta)&=\sum_{i=1}^{n}\sum_{j=1}^{p_{xn}}(Y_i-\hat{d}_{ij}\beta_j)^2+\lambda_n^*\sum_{j=1}^{p_{xn}}|\beta_j|^{\gamma}\\
   & =\sum_{j=1}^{p_{xn}}\left[\sum_{i=1}^{n}(Y_i-\hat{d}_{ij}\beta_j)^2\right]+\lambda_n^*\sum_{j=1}^{p_{xn}}|\beta_j|^{\gamma}\\
   & =\sum_{j=1}^{p_{xn}}\left[\sum_{i=1}^{n}(Y_i\pm d_{1i}'\beta\pm d_{ij}\beta_j-\hat{d}_{ij}\beta_j)^2\right]+\lambda_n^*\sum_{j=1}^{p_{xn}}|\beta_j|^{\gamma}\\
    & =\sum_{j=1}^{p_{xn}}\left[\sum_{i=1}^{n}(Y_i- d_{1i}'\beta+d_{1i}'\beta- d_{ij}\beta_j+(d_{ij}-\hat{d}_{ij})\beta_j)^2\right]+\lambda_n^*\sum_{j=1}^{p_{xn}}|\beta_j|^{\gamma}\\
    &  =\sum_{j=1}^{p_{xn}}\left[\sum_{i=1}^{n}(\varepsilon_i+d_{1i}'\beta- d_{ij}\beta_j+(d_{ij}-\hat{d}_{ij})\beta_j)^2\right]+\lambda_n^*\sum_{j=1}^{p_{xn}}|\beta_j|^{\gamma}\\
    &  =\sum_{j=1}^{p_{xn}}\left[\sum_{i=1}^{n}[\varepsilon_i^2+2\varepsilon_i(d_{1i}'\beta- d_{ij}\beta_j+(d_{ij}-\hat{d}_{ij})\beta_j)+(d_{1i}'\beta- d_{ij}\beta_j+(d_{ij}-\hat{d}_{ij})\beta_j)^2]\right]+\lambda_n^*\sum_{j=1}^{p_{xn}}|\beta_j|^{\gamma}\\
     &  =\sum_{j=1}^{p_{xn}}\left[\sum_{i=1}^{n}[\varepsilon_i^2+2\varepsilon_i(d_{1i}'\beta- d_{ij}\beta_j+(d_{ij}-\hat{d}_{ij})\beta_j)+(d_{1i}'\beta- d_{ij}\beta_j)^2+2(d_{1i}'\beta- d_{ij}\beta_j)(d_{ij}-\hat{d}_{ij})\beta_j+(d_{ij}-\hat{d}_{ij})^2\beta_j^2]\right]+\lambda_n^*\sum_{j=1}^{p_{xn}}|\beta_j|^{\gamma}\\
     &  =\sum_{j=1}^{p_{xn}}\left[\sum_{i=1}^{n}[\varepsilon_i^2+2\varepsilon_i(d_{1i}'\beta- d_{ij}\beta_j)+(d_{1i}'\beta- d_{ij}\beta_j)^2\right.\\
     &\left.-2\varepsilon_i((d_{ij}-\hat{d}_{ij})\beta_j)+2(d_{1i}'\beta- d_{ij}\beta_j)(d_{ij}-\hat{d}_{ij})\beta_j+(d_{ij}-\hat{d}_{ij})^2\beta_j^2]\right]+\lambda_n^*\sum_{j=1}^{p_{xn}}|\beta_j|^{\gamma}\\
          &  =\sum_{j=1}^{p_{xn}}\left[\sum_{i=1}^{n}[\varepsilon_i^2-2\varepsilon_i d_{ij}\beta_j- 2d_i\beta_{10}d_{ij}\beta_j+(d_{1i}'\beta)^2+ d_{ij}\beta_j)^2+2\varepsilon_id_{1i}'\beta\right.\\
     &\left.-2\varepsilon_i((d_{ij}-\hat{d}_{ij})\beta_j)+2(d_{1i}'\beta- d_{ij}\beta_j)(d_{ij}-\hat{d}_{ij})\beta_j+(d_{ij}-\hat{d}_{ij})^2\beta_j^2]\right]+\lambda_n^*\sum_{j=1}^{p_{xn}}|\beta_j|^{\gamma}\\
     &  =\sum_{j=1}^{p_{xn}}\left[\sum_{i=1}^{n}\varepsilon_i^2-2(\varepsilon_n a_j+n\xi_{nj})\beta_j+ \sum_{i=1}^{n}d_{ij}^2\beta_j^2 +\sum_{i=1}^{n}y_i^2-\sum_{i=1}^{n}\varepsilon_i^2 \right.\\
     &\left.-2\sum_{i=1}^{n}\varepsilon_i((d_{ij}-\hat{d}_{ij})\beta_j)+2\sum_{i=1}^{n}(d_{1i}'\beta- d_{ij}\beta_j)(d_{ij}-\hat{d}_{ij})\beta_j+\sum_{i=1}^{n}(d_{ij}-\hat{d}_{ij})^2\beta_j^2]\right]+\lambda_n^*\sum_{j=1}^{p_{xn}}|\beta_j|^{\gamma}\\
\end{align*}

So, minimizing $U^*(\beta)$ is equivalent to minimize:
\begin{align*}
    &\sum_{j=1}^{p_{xn}}\left[d_{nj}'d_{nj}\beta_j^2-2(\varepsilon_n'a_j+n\xi_{nj})\beta_j+(d_{nj}-\hat{d}_{nj})'(d_{nj}-\hat{d}_{nj})\beta_j^2\right.\\
    &\left. -2\varepsilon_n'(d_{nj}-\hat{d}_{nj})\beta_j+2(D_{1n}'\beta_{10}-d_{nj}\beta_{j})'(d_{nj}-\hat{d}_{nj})\beta_j\right]+\lambda_{xn}^*\sum_{j=1}^{p_{xn}}|\beta_j|^{\gamma}\\
    &=\sum_{j=1}^{p_{xn}}\left[(d_{nj}'d_{nj}+(d_{nj}-\hat{d}_{nj})'(d_{nj}-\hat{d}_{nj}))\beta_j^2-2((\varepsilon_n'a_j+n\xi_{nj})+\right.\\ 
    &\left. \varepsilon_n'(d_{nj}-\hat{d}_{nj})-(D_{1n}'\beta_{10}-d_{nj}\beta_{j})'(d_{nj}-\hat{d}_{nj}))\beta_j\right]+\lambda_{xn}^*\sum_{j=1}^{p_{xn}}|\beta_j|^{\gamma}\\
\end{align*}

Let $g(\beta_j^*)=c_n\beta_j^2-2(\varepsilon_n'a_j+n\xi_{nj}+r_n)+\lambda_n^*|\beta_j|^{\gamma}$, where $c_n= d_{nj}'d_{nj}+(d_{nj}-\hat{d}_{nj})'(d_{nj}-\hat{d}_{nj})= nO_p(1)+O_p(1)=O_p(n)$ by the instruments being $O_p(1)$ and the true 1st stage coefficients being uniformly bounded. Note that at least the 1st term is also bounded away from 0, due to variation in the instruments and existence of 1st stage coefficients that are bounded away from 0. If all $\alpha$ was identically 0, $E[XZ]= 0$ which contradicts the premise. Also, $r_n=\varepsilon_n'(d_{nj}-\hat{d}_{nj})-(D_{1n}'\beta_{10}-d_{nj}\beta_{j})'(d_{nj}-\hat{d}_{nj}) =n^{1/2}O_p(1)+k_nn^{1/2}O_p(1)=O_p(k_nn^{1/2})$.

Then, by Lemma \ref{lemma:KnightFu} and the analysis above, $\beta_j=0$ is the solution to $\argmin g(\beta_j)$ iff $c_n^{-1}\lambda_{xn}^*>c_{\gamma}|c_n^{-1}a_n|^{2-\gamma}=c_\gamma(c_n^{-1}|\varepsilon_n'a_j+n\xi_{nj}+r_n|)^{2-\gamma}$.  Expanding on that:

\begin{align*}
    &c_n^{-1}\lambda_{xn}^*>c_\gamma(n^{-1}O_p(1)|\varepsilon_n'a_j+n\xi_{nj}+r_n|)^{2-\gamma}\\
\Rightarrow  & n^{-1}O_p(1)c_{\gamma}^{-1}\lambda_{xn}^*>n^{-(2-\gamma)}O_p(1)|\varepsilon_n'a_j+n\xi_{nj}+r_n|^{2-\gamma}\\
\Rightarrow & n^{-\frac{1}{2-\gamma}}c_{\gamma}^{-\frac{1}{2-\gamma}}\lambda_{xn}^{*\frac{1}{2-\gamma}}O_p(1)>|\varepsilon_n'a_j+n\xi_{nj}+r_n|\\
\Rightarrow & (\frac{\lambda_{xn}^*}{n^{\gamma/2}})^{\frac{1}{2-\gamma}}c_{\gamma}^{-\frac{1}{2-\gamma}}O_p(1)>n^{-1/2}|\varepsilon_n'a_j+n\xi_{nj}+r_n|\\
\end{align*}
Let $w_n=c_{\gamma}^{-\frac{1}{2-\gamma}}(\lambda/n^{\gamma/2})^{\frac{1}{2-\gamma}}O_p(1)$. Suffices to show that:
\begin{enumerate}[i]
    \item $P(w_n>n^{-1/2}\max_{j\in J_n}|\varepsilon_n'a_j+n\xi_{nj}+r_n|)\to1$.
    \item $P(w_n>n^{-1/2}\min_{j\in K_n}|\varepsilon_n'a_j+n\xi_{nj}+r_n|)\to0$.
\end{enumerate}

For (i): $P(w_n>n^{-1/2}\max_{j\in J_n}|\varepsilon_n'a_j+n\xi_{nj}+r_n|)\to1$.
By the partial orthogonality assumption (C.2a), $\exists c_0>0$ st. \[\left|n^{-1/2}\sum_{i=1}^{n}d_{ij}d_{ik}\right|\leq c_0,\ j=1,\dots, m_{xn},\ k=1,\dots, k_{xn}, \text{for $n$ large enough}.\]

Therefore, $n^{1/2}|\xi_{nj}|=n^{1/2}\left|n^{-1}\sum_{i=1}^{n}(d_{1i}'\beta_{10})d_{ij}\right|\leq n^{-1/2}b_1\sum_{l=1}^{k_n}\left|\sum_{i=1}^{n}d_{il}d_{ij}\right|\leq b_1k_nc_0\leq c_1k_n$ for every $j\in J_n$ and $c_1=c_0b_1$. The first inequality holds by (C.4).

Then, \begin{align*}
    &P(w_n>n^{-1/2}\max_{j\in J_n}|\varepsilon_n'a_j+n\xi_{nj}+r_n|)\\
    =&P(w_n>n^{-1/2}\max_{j\in J_n}|\varepsilon_n'a_j+n\xi_{nj}+O_p(k_nn^{1/2})|)\\
    \geq &P(w_n>n^{-1/2}\max_{j\in J_n}|\varepsilon_n'a_j|+c_1k_n+k_nO_p(1))\\
    =&1- P(n^{-1/2}\max_{j\in J_n}|\varepsilon_n'a_j|\leq w_n-c_1k_n-k_nO_p(1))\\
    \geq &1-\frac{E[n^{-1/2}\max_{j\in J_n}|\varepsilon_n'a_j|]}{w_n-c_1k_n-k_nO_p(1)}=1- \frac{(log(2))^{1/2}K(log(m_{xn})^{1/2})}{w_n-c_1k_n-k_nO_p(1)}
\end{align*}

The numerator is sub-Gaussian by Lemma 4 in Huang et al. (2008,), hence the last equation. Note that, 
\[\frac{k_n(c_1+O_p(1))}{c_{\gamma}^{-\frac{1}{2-\gamma}}(\lambda/n^{\gamma/2})^{1/(2-\gamma)}O_p(1)}=\left(\frac{k_nc_1}{c_{\gamma}(\lambda/n^{\gamma/2})O_p(1)}\right)^{1/(2-\gamma)}+\frac{k_nO_p(1)}{c_{\gamma}^{-\frac{1}{2-\gamma}}(\lambda/n^{\gamma/2})^{1/(2-\gamma)}O_p(1)}=\]\[ \left(\frac{k_n}{c_{\gamma}(\lambda/n^{\gamma/2})O_p(1)}\right)^{1/(2-\gamma)}+\left(\frac{k_nO_p(1)}{c_{\gamma}(\lambda/n^{\gamma/2})O_p(1)}\right)^{1/(2-\gamma)}\to 0\] which is $o_p(1)$ by assumption, so the leading term is of order $\frac{log(m_{xn})}{(\lambda/n^{\gamma/2})^{\frac{2}{2-\gamma}}O_p(1)}$ which is $o_p(1)$ by (C.3b).

For (ii),  I need to show $P(w_n>n^{-1/2}\min_{j\in K_n}|\varepsilon_n'a_j+n\xi_{nj}+r_n|)\to0$, ie.

\begin{align*}
    &P(w_n>n^{-1/2}\min_{j\in K_n}|\varepsilon_n'a_j+n\xi_{nj}+r_n|)\\
    = &P(w_n>n^{-1/2}\min_{j\in K_n}|\varepsilon_n'a_j+n\xi_{nj}+O_p(k_nn^{1/2})|)\\
   = &P\left(\bigcup_{1\leq j\leq k_{xn}} n^{-1/2}|\varepsilon_n'a_j+n\xi_{nj}+O_p(k_nn^{1/2})|< w_n\right)\\
    \leq &\sum_{j=1}^{k_{xn}}P\left(n^{-1/2}|\varepsilon_n'a_j+n\xi_{nj}+O_p(k_nn^{1/2})|< w_n\right)\\
\end{align*}

For each term, I also have: \begin{align*}
   & P\left(n^{-1/2}|\varepsilon_n'a_j+n\xi_{nj}+O_p(k_{xn}n^{1/2})|< w_n\right)\\
   =& 1- P\left(n^{-1/2}|\varepsilon_n'a_j+n\xi_{nj}+O_p(k_{xn}n^{1/2})|>w_n\right)\\
   \leq& 1- P\left(n^{1/2}|\xi_{nj}|+O_p(k_{xn})-n^{-1/2}|\varepsilon_n'a_j|>w_n\right)\\
   =& P\left(n^{-1/2}|\varepsilon_n'a_j|>n^{1/2}|\xi_{nj}|+O_p(k_{xn})-w_n\right)\\
   \leq & Kexp[-C(n^{1/2}\xi_0-w_n+k_{xn}O_p(1))^2]
\end{align*}

Then, 
\begin{align*}
    P(w_n>n^{-1/2}\min_{j\in K_n}|\varepsilon_n'a_j+n\xi_{nj}+r_n|)&\leq k_{xn}Kexp[-C(n^{1/2}\xi_0-w_n+k_{xn}O_p(1))^2]\\
    &\leq k_{xn}Kexp[-Cn(\xi_0-\frac{w_n}{n^{1/2}}+\frac{k_{xn}}{n^{1/2}}O_p(1))^2]\\
    &\leq k_{xn} K exp(-Cn)\leq O_p(1)\frac{k_{xn}}{n}=o_p(1).
\end{align*}
where the 3rd inequality holds by:
\[\frac{k_{xn}w_n}{n^{1/2}}=\frac{k_{xn}c_{\gamma}^{-\frac{1}{2-\gamma}}(\lambda_{xn}/n^{\gamma/2})^{\frac{1}{2-\gamma}}O_p(1)}{n^{1/2}}=O_p(1)\left(\frac{k_{xn}^{\gamma-2}\lambda_{xn}n^{-\gamma/2}}{n^{\frac{2-\gamma}{2}}}\right)^{\frac{1}{2-\gamma}}=O_p(1)(\frac{k_{xn}^{\gamma-2}\lambda_{xn}}{n})^{\frac{1}{2-\gamma}}\] From Assumption (C.3) $\lambda_nn^{-\gamma/2}k_n^{\gamma-2}\to \infty$ and $\lambda_{xn}/n\to 0$, so $\lambda_{xn}=o_p(n)$. Thus, $n*n^{-\gamma/2}k_{xn}^{\gamma-2}=n^{(2-\gamma)/2}k_{xn}^{\gamma-2}=\left(\frac{n^{1/2}}{k_{xn}}\right)^{2-\gamma}\to \infty$, which implies that $k_{xn}/n^{1/2}\to 0$. Which concludes the proof.

\textbf{Proof of Theorem \ref{thm:large_oracle_bridge}}\\
The proof of this theorem is a direct application of the theorem \ref{thm:oracle} and the corresponding theorem in \citeN{Huang:2008}. By sparsity, $k_{xn}<n$, and I have already proven model selection consistency, so I apply the ``$p<n$" section for $k_{xn}$.
\newpage
 \begin{proposition}
\label{thm:proposition_ad}
    Define $\hat{\beta}_n=_s\beta$ if and only if $sgn(\hat{\beta}_n)=sgn(\beta)$. Let $W_{n1}=diag(w_{n1},\dots,w_{nk_{xn}})$, $w_{n1}=\{w_{n1},\dots,w_{nk_{xn}}\}$, $W_{n2}=diag(w_{nk_{xn}+1},\dots,w_{np_{xn}})$, $w_{n2}=\{w_{nk_{xn}+1},\dots,w_{np_{xn}}\}$. Then, \[P(\hat{\beta}_n=_s\beta_0)\geq P(A_n\cap B_n),\] where \[A_n=\{2n^{-1/2}|(\hat\Sigma_{n11})^{-1}\hat D_{1n}'\varepsilon_{n}|\leq 2\sqrt{n}|\beta_{10}|-n^{-1/2}\lambda_{xn}|(\hat\Sigma_{n1})^{-1}W_{n1}sgn(\beta_{10})|\}\] and \[B_n= \{|2n^{-1/2}\hat D_{2n}'(1-\hat H )\varepsilon_n|\leq n^{-1/2}\lambda_{xn}w_{n2}-n^{-1/2}\lambda_{xn}|\hat\Sigma_{n21}(\hat\Sigma_{n11})^{-1}W_{n1}sgn(\beta_{10})|\}\]
     
 \end{proposition}
\textbf{Proof:} The proof of the proposition follows the same steps with \cite{Zhao:2006} by adjusting the optimization problem accordingly.

\textbf{Proof of Theorem \ref{thm:large_msc_ad}}

Under proposition , $P(\hat{\beta}_n=_s\beta_0)>1-P(A_n^c)-P(B_n^c)$, so it suffices to show that the latter two events happen with probability converging to $0$. 

For $A_n^c$, let $\eta_n=2n^{-1/2}(\Sigma_{1}^d)^{-1}D_{1n}\varepsilon_n$, $u_n=(\hat\Sigma_{1}^d)^{-1}W_{n1}sgn(\beta_{10})$.

\begin{align*}
    2n^{-1/2}|(\hat\Sigma_{1}^d)^{-1}\hat D_{1n}\varepsilon_n|&\leq 2n^{-1/2}|(\hat\Sigma_{1}^d)^{-1}D_{1n}\varepsilon_n|+2n^{-1/2}|(\hat \Sigma_{1}^d)^{-1}(\hat D_{1n}-D_{1n})\varepsilon_n|\\
    &\leq 2n^{-1/2}|(\Sigma_{1}^d)^{-1}D_{1n}\varepsilon_n|+2n^{-1/2}|((\Sigma_{1}^d)^{-1}-(\hat\Sigma_{1}^d)^{-1})D_{1n}\varepsilon_n|\\
    &+2n^{-1/2}|(\hat \Sigma_{1}^d)^{-1}(\hat D_{1n}-D_{1n})\varepsilon_n|\\
    &\leq |\eta_n|+o_p(1)
\end{align*}
where the last inequality holds by Chebyshev and (D6.a).

The proof for the 2nd term being $o_p(1)$ goes as follows:
Define \[T_n=n^{-1/2}((\Sigma_{1}^d)^{-1}-(\hat\Sigma_{1}^d)^{-1})D_{1n}\varepsilon_n=n^{-1/2}\sum_{i=1}^{n}\varepsilon_i\delta_n'((\Sigma_{1}^d)^{-1}-(\hat\Sigma_{1}^d)^{-1})d_{1i}=n^{-1/2}\sum_{i=1}^{n}\varepsilon_iu_i\] where $\norm{\delta_n}=1$.
Note that $E(T_n|u_n)=0$ and $Var(T_n|u_n)=n^{-1}\sigma_{\varepsilon}^2\sum_{i=1}^{n}u_{i}^2$.
Also, \begin{align*}
    n^{-1}\sum_{i=1}^{n}u_{i}^2&=n^{-1}\sum_{i=1}^{n}(\delta_n'((\Sigma_{1}^d)^{-1}-(\hat\Sigma_{1}^d)^{-1})d_{1i})^2\\
    &= \delta_n'((\Sigma_{1}^d)^{-1}-(\hat\Sigma_{1}^d)^{-1})\left(\frac{1}{n}\sum_{i=1}^{n}d_{1i}'d_{1i}\right)\delta_n'((\Sigma_{1}^d)^{-1}-(\hat\Sigma_{1}^d)^{-1})\\
    &\leq \norm{(\Sigma_{1}^d)^{-1}-(\hat\Sigma_{1}^d)^{-1}}^2\norm{\frac{1}{n}\sum_{i=1}^{n}d_{1i}'d_{1i}}^2\leq \norm{(\Sigma_{1}^d)^{-1}}^2\norm{(\hat\Sigma_{1}^d)^{-1}}^2\norm{\Sigma_{1}^d-\hat\Sigma_{1}^d}^2\norm{\frac{1}{n}\sum_{i=1}^{n}d_{1i}'d_{1i}}^2\\
    &\leq (\tau_{1n}^{d}+o_p(1))^{-2}({\tau}_{1n}^{d})^{-2}\frac{k_{xn}\sqrt{\max_jk_{znj}}}{\sqrt{n}}O_p(1)\tau_{2d}^{2}=o_p(1)
\end{align*}
where the last inequality follows by the consistency of the eigenvalues due to Weyl's theorem, the fact that they are bounded away from and the relevant part of the proof of theorem \ref{thm:oracle}.  So, $\forall \epsilon$, I can find $\delta$,
\begin{align*}
    P(|T_n|>\epsilon)&=P(\{|T_n|>\epsilon\}\cap\{|n^{-1}\sum_{i=1}^{n}u_i^2|>\mu\})+P(\{|T_n|>\epsilon\}\cap\{|n^{-1}\sum_{i=1}^{n}u_i^2|\leq\mu\})\\
    &\leq P(\{|T_n|>\epsilon\}\cap\{|n^{-1}\sum_{i=1}^{n}u_i^2|>\mu\})+P(\{|n^{-1}\sum_{i=1}^{n}u_i^2|\leq\mu\})\\
    &\leq \frac{\sigma_{\varepsilon}^2\mu}{n\epsilon^2}+\frac{\delta}{2}\leq \frac{\delta}{2}+\frac{\delta}{2}=\delta.
\end{align*}
The proof for the last term is part of the proof of theorem \ref{thm:large_msc} and (D.6a). By the same arguments, 
\begin{align*}
    2n^{-1/2}\norm{(\hat\Sigma_{1}^d)^{-1}\hat D_{1n}\varepsilon_n-(\Sigma_{1}^d)^{-1}D_{1n}\varepsilon_n}&= 2n^{-1/2}\norm{((\hat\Sigma_{1}^d)^{-1}-(\Sigma_{1}^d)^{-1})D_{1n}\varepsilon_n+(\hat \Sigma_{1}^d)^{-1}(\hat D_{1n}-D_{1n})\varepsilon_n}\\
    &\leq 2n^{-1/2}\norm{((\hat\Sigma_{1}^d)^{-1}-(\Sigma_{1}^d)^{-1})D_{1n}\varepsilon_n}\\
    &+ 2n^{-1/2}\norm{(\hat \Sigma_{1}^d)^{-1}(\hat D_{1n}-D_{1n})\varepsilon_n}\\
    &\leq o_p(1)
\end{align*}


Note that 
\begin{align*}
    n^{-1/2}\lambda_{xn}|(\hat\Sigma_{n1})^{-1}W_{n1}sgn(\beta_{10})|&\leq n^{-1/2}\lambda_{xn}|((\Sigma_{n1})^{-1}-(\hat\Sigma_{n1})^{-1})W_{n1}sgn(\beta_{10})|\\
    &+n^{-1/2}\lambda_{xn}|(\Sigma_{n1})^{-1}W_{n1}sgn(\beta_{10})|\\
    &\leq n^{-1/2}\lambda_{xn}\norm{(\Sigma_{n1})^{-1}-(\hat\Sigma_{n1})^{-1}}\norm {W_{n1}}+ n^{-1/2}\lambda_{xn}|(\Sigma_{n1})^{-1}W_{n1}sgn(\beta_{10})|\\
    &\leq \frac{\lambda_{xn}\sqrt{k_{xn}}}{\sqrt{n}}o_p(1)+ n^{-1/2}\lambda_{xn}|(\Sigma_{n1})^{-1}W_{n1}sgn(\beta_{10})|\\
    &\leq  n^{-1/2}\lambda_{xn}|(\Sigma_{n1})^{-1}W_{n1}sgn(\beta_{10})|+o_p(1)
\end{align*}

where the last two inequalities hold by assumptions (D2.b) and (D6.a) respectively. By similar arguments, 

\begin{align*}
    &n^{-1/2}\lambda_{xn}\norm{(\hat\Sigma_{n1})^{-1}W_{n1}sgn(\beta_{10})-(\Sigma_{n1})^{-1}W_{n1}sgn(\beta_{10})}\\
    + &n^{-1/2}\lambda_{xn}\norm{((\Sigma_{n1})^{-1}-(\hat\Sigma_{n1})^{-1})W_{n1}sgn(\beta_{10})} \leq o_p(1)
\end{align*}


\begin{align*}
    A_n'&= \{2n^{-1/2}|(\Sigma_{n1})^{-1} D_{1n}'\varepsilon_{n}|+o_p(1)\leq 2\sqrt{n}|\beta_{10}|-n^{-1/2}\lambda_{xn}|(\Sigma_{n1})^{-1}W_{n1}sgn(\beta_{10})|-o_p(1)\}\\
    &=\{2n^{-1/2}|(\Sigma_{n1})^{-1} D_{1n}'\varepsilon_{n}|\leq 2\sqrt{n}|\beta_{10}|-n^{-1/2}\lambda_{xn}|(\Sigma_{n1})^{-1}W_{n1}sgn(\beta_{10})|+o_p(1)\},
\end{align*}
Given the continuity of the distribution of the error and the consistency of both sides of the inequality in $A_n$ to the corresponding quantities in $A_n'$, it suffices to show that  $P(A_n^{'c})\to 0$. The rest of the proof for the first half goes as in \cite{Huang:2008b}. That is, I can decompose the event of interest as such:

\begin{align*}
    P(A_n^{'c})&=P(A_n^{'c}\cap\{|w_{n1}|\leq c_1 b_{1}^{-1}\})+P(A_n^{'c}\cap\{|w_{n1}|> c_1 b_{1}^{-1}\})\\
    &=P(A_n^{'c}\cap\{|w_{n1}|\leq c_1 b_{1}^{-1}\})+P(|w_{n1}|> c_1 b_{1}^{-1})
\end{align*}
where $c_1$ is a constant and the second term is very small from:
\begin{align*}
    P(\min_{1\leq j\leq k_{xn}}|\tilde\beta_{nj}|\geq c_1^{-1} b_{1})>1-\epsilon & \Rightarrow P(\min_{1\leq j\leq k_{xn}}w_{nj}\geq c_1 b_{1}^{-1})\leq \epsilon\\
    &\Rightarrow P(w_{nj}\geq c_1 b_{1}^{-1})\leq \epsilon
\end{align*}

by the consistency of the initial estimator.

So, it suffices to show that $P(A_n^{'c}\cap\{|w_{n1}|\leq c_1 b_{1}^{-1}\})\to 0$. By the spectral decomposition: $(\Sigma_{1n}^{d})^{-1}=\sum_{j=1}^{k_{xn}}(\tau_{nj}^d)^{-1}\gamma_j\gamma_j'$ where $\tau_{nj}^d$ is the j-th eigenvalue and $\gamma_j$ the corresponding eigenvector.

Define $u_n= (\Sigma_{1n}^{d})^{-1}(W_{n1}sgn(\beta_{10}))'=\sum_{j=1}^{k_{xn}}(\tau_{nj}^d)^{-1}\gamma_j\gamma_j'(W_{n1}sgn(\beta_{10}))' $. Then, by Cauchy-Schwartz, for every element $1\leq l\leq k_{xn}$, $\norm{u_l}^2\leq \tau_{1}^{-2}\sum_{j=1}^{k_{xn}}\norm{\gamma_j}^2\norm{w_{n1}}^2\leq \tau_{1}^{-2}c_1b_{1}^{-2}k_{xn}^2$. Thus, $|u_l|\leq c_1'b_{1}^{-1}k_{xn}$.

Let $v_n=2\sqrt{n}b_1-c_1'n^{-1/2}\lambda_{xn}k_{xn}b_{1}^{-1}$. Define the event $C_{n1}:=\{|\eta_j|\geq v_n+o_{p}(1),1\leq j \leq k_{xn}\}=\{\max_{1\leq j \leq k_{xn}}|\eta_j|\geq v_{n}+o_p(1)\}$.

By Lemma 1 in \cite{Huang:2008b} and $\eta_j$ being sub-Gaussian:
\[P\left(\max_{1\leq j \leq k_{xn}}|\eta_j|\geq v_{n}+o_p(1)\right)\leq \frac{K'(\log k_{xn})^{1/d}}{v_n+o_p(1)}\] for a constant $K'=(\log 2)^{1/2}K$. Then, for $d\in [1,2]$, by (D4), the probability of this event goes to 0. Then, $P(A_n^{'c}\cap\{|w_{n1}|\leq c_1b_1^{-1}\})\to 0$.

Similarly for $B_n^c$, note that:

\begin{align*}
    n^{-1/2}|\hat D_{2n}'(1-\hat H )\varepsilon_n|&\leq n^{-1/2}| D_{2n}'(1-\hat H )\varepsilon_n|+n^{-1/2}|(\hat D_{2n}- D_{2n})'(1-\hat H )\varepsilon_n|\\
    &\leq  n^{-1/2}| D_{2n}'(1- H )\varepsilon_n|+n^{-1/2}|(\hat D_{2n}- D_{2n})'(1-\hat H )\varepsilon_n|\\
    &+ n^{-1/2}| D_{2n}'(\hat{H}-H)\varepsilon_n|\\
   & \leq n^{-1/2}| D_{2n}'(1- H )\varepsilon_n|+o_p(1)
\end{align*}

where both residual terms are $o_p(1)$ by Chebyshev's inequality:
Define $T_n =n^{-1/2}\sum_{i=1}^{n}\varepsilon_i\delta_n'(1-\hat H)(\hat d_{2i}-d_{2i})=n^{-1/2}\sum_{i=1}^{n} \varepsilon_iu_i$. Note that $E(T_n|u_n)=0$, and $Var(T_n|u_n)=\frac{\sigma_{\varepsilon}^{2}}{n}\sum_{i=1}^{n}u_i$, where the latter is equal to:
\begin{align*}
    \frac{1}{n}\sum_{i=1}^{n}u_i^2&=\delta_n'(1-\hat H)\Delta_2'\left(\frac{1}{n}\sum_{i=1}^{n}Z_i'Z_i\right)_2\Delta(1-\hat H)\delta_n\\
    &\leq \norm{1-\hat H}^2\norm{\Delta_2}^2\norm{\frac{1}{n}\sum_{i=1}^{n}Z_i'Z_i}\leq 2\tau_{2z}\frac{k_{xn}\max_j k_{znj}}{n}O_p(1)=o_p(1)
\end{align*}
because $\norm{\hat H}=\frac{1}{n}\norm{\hat{D}_{n1}'(\hat\Sigma_{1n}^{d})^{-1}\hat{D}_{n1}}\leq \frac{1}{n}\norm{\hat{D}_{n1}}^2\norm{(\hat\Sigma_{1n}^{d})^{-1}}= \norm{(\hat\Sigma_{1n}^{d})}\norm{(\hat\Sigma_{1n}^{d})^{-1}}=1$
Then, by Chebyshev and the last sum being $o_p(1)$, the whole term is $o_p(1)$.
The argument for the 2nd residual form is exactly the same, by noticing that:
\begin{align*}
    \norm{\hat H- H}&= \norm{n^{-1}\hat{D}_{n1}'(\hat\Sigma_{1n}^{d})^{-1}\hat{D}_{n1}-n^{-1}D_{n1}'(\Sigma_{1n}^{d})^{-1}{D}_{n1}}\\
    &\leq \norm{n^{-1}\hat{D}_{n1}'((\hat\Sigma_{1n}^{d})^{-1}-(\Sigma_{1n}^{d})^{-1})\hat{D}_{n1}}+\norm{(\hat{D}_{n1}\pm D_{n1})'(\Sigma_{1n}^{d})^{-1}\hat{D}_{n1}-D_{n1}'(\Sigma_{1n}^{d})^{-1}D_{n1}}\\
    &\leq \norm{n^{-1}\hat{D}_{n1}'\hat{D}_{n1}}\norm{(\hat\Sigma_{1n}^{d})^{-1}-(\Sigma_{1n}^{d})^{-1}}+\\
    &+\norm{n^{-1}(\hat{D}_{n1}- D_{n1})'(\Sigma_{1n}^{d})^{-1}\hat{D}_{n1}+n^{-1}D_{n1}'(\Sigma_{1n}^{d})^{-1}(\hat D_{n1}\pm D_{n1})-n^{-1}D_{n1}'(\Sigma_{1n}^{d})^{-1}D_{n1}}\\
    &\leq \hat{\tau}_{2n}^d    \norm{(\hat\Sigma_{1n}^{d})^{-1}-(\Sigma_{1n}^{d})^{-1}}+\norm{n^{-1}(\hat{D}_{n1}- D_{n1})'(\Sigma_{1n}^{d})^{-1}\hat{D}_{n1}}\\
    &+\norm{n^{-1}D_{n1}'(\Sigma_{1n}^{d})^{-1}(\hat D_{n1}- D_{n1})}\\
    &\leq (\tau_{2}^d+o_p(1))\frac{k_{xn}\sqrt{\max_j k_{znj}}}{\sqrt{n}}O_p(1)+n^{-1/2}n^{1/2}(\tau_2^d+o_p(1))(\tau_2^d)^{-1}\norm{\hat D_{n1}- D_{n1}}\\
    &+n^{-1/2}n^{1/2}\tau_2^d((\tau_2^d)^{-1}+o_p(1))\norm{\hat D_{n1}- D_{n1}}\\
    &\leq o_p(1)+O_p(1)\norm{\hat D_{n1}- D_{n1}}=o_p(1)
\end{align*}
Also,
\begin{align*}
   & n^{-1/2}\norm{\hat D_{2n}'(1-\hat H )\varepsilon_n- D_{2n}'(1- H )\varepsilon_n}\\
    \leq&  n^{-1/2}\norm{(\hat D_{2n}- D_{2n})'(1-\hat H )\varepsilon_n+ D_{2n}'(\hat{H}-H)\varepsilon_n} \leq o_p(1)
\end{align*}
For the other side of the inequality in $B_n$, note that 
\begin{align*}
    &n^{-1/2}\lambda_{xn}|(\hat\Sigma_{n21})^{-1}(\hat\Sigma_{n1})^{-1}W_{n1}sgn(\beta_{10})|\\
    \leq & n^{-1/2}\lambda_{xn}|(\hat\Sigma_{n21})^{-1}((\Sigma_{n1})^{-1}-(\hat\Sigma_{n1})^{-1})W_{n1}sgn(\beta_{10})|+n^{-1/2}\lambda_{xn}|(\hat\Sigma_{n21})^{-1}(\Sigma_{n1})^{-1}W_{n1}sgn(\beta_{10})|\\
    \leq & n^{-1/2}\lambda_{xn}|(\hat\Sigma_{n21})^{-1}((\Sigma_{n1})^{-1}-(\hat\Sigma_{n1})^{-1})W_{n1}sgn(\beta_{10})|\\
    +&n^{-1/2}\lambda_{xn}|(\hat\Sigma_{n1})^{-1}((\Sigma_{n21})^{-1}-(\hat\Sigma_{n21})^{-1})W_{n1}sgn(\beta_{10})|+n^{-1/2}\lambda_{xn}|(\Sigma_{n21})^{-1}(\Sigma_{n1})^{-1}W_{n1}sgn(\beta_{10})|\\
    \leq  & n^{-1/2}\lambda_{xn}\norm{(\hat\Sigma_{n21})^{-1}}\norm{(\Sigma_{n1})^{-1}-(\hat\Sigma_{n1})^{-1}}\norm {W_{n1}}+n^{-1/2}\lambda_{xn}\norm{(\hat\Sigma_{n1})^{-1}}\norm{(\Sigma_{n21})^{-1}-(\hat\Sigma_{n21})^{-1}}\norm {W_{n1}}\\
    +& n^{-1/2}\lambda_{xn}|(\Sigma_{n21})^{-1}(\Sigma_{n1})^{-1}W_{n1}sgn(\beta_{10})|\\
    \leq& \frac{\lambda_{xn}\sqrt{k_{xn}}}{\sqrt{n}}\rho_{1d}^{-1}o_p(1)+ n^{-1/2}\lambda_{xn}\tau_{1d}^{-1}\rho_{2d}p_{xn}(k_{xn}\max_jk_{znj}/n)^{1/2}+ n^{-1/2}\lambda_{xn}|(\Sigma_{n21})^{-1}(\Sigma_{n1})^{-1}W_{n1}sgn(\beta_{10})|\\
    \leq & n^{-1/2}\lambda_{xn}|(\Sigma_{n21})^{-1}(\Sigma_{n1})^{-1}W_{n1}sgn(\beta_{10})|+o_p(1) 
\end{align*}

By similar steps, \[n^{-1/2}\lambda_{xn}\norm{(\hat\Sigma_{n21})^{-1}(\hat\Sigma_{n1})^{-1}W_{n1}sgn(\beta_{10}) - (\Sigma_{n21})^{-1}(\Sigma_{n1})^{-1}W_{n1}sgn(\beta_{10})}=o_p(1).\]

The rest of the proof for the second event happening with probability $0$, follows the proof from the online appendix of \citeN{Huang:2008b}, using the rationale on the first part of this proof to get away with the extra $o_p(1)$ term and using the fact that the $Z_i$ is $O_p(1)$.

\textbf{Proof of Theorem \ref{thm:large_oracle_ad}}\\
The proof of this theorem follows similar steps in verifying the Lidemberg conditions as in theorem \ref{thm:oracle} and the corresponding theorem in \citeN{Huang:2008}.

\end{appendices}

\end{document}